\documentclass[twocolumn,twocolappendix]{aastex631}

\usepackage{natbib}
\usepackage{graphicx}	
\usepackage{amsmath}	
\usepackage{amssymb}	
\usepackage{bm}	
\usepackage{subfigure}
\usepackage{longtable}
\usepackage{array}
\usepackage{booktabs}
\usepackage{multirow}
\usepackage{xcolor}
\usepackage{lipsum}

\shorttitle{GRB Jets in CSM}
\shortauthors{Hamidani, et al.}

\begin{document}
\title{Gamma-Ray Burst Jets in Circumstellar Material: Dynamics, Breakout, and Diversity of Transients}

\correspondingauthor{Hamid Hamidani}
\email{hhamidani@astr.tohoku.ac.jp}

\author[0000-0003-2866-4522]{Hamid Hamidani}
\affiliation{Astronomical Institute, Graduate School of Science, Tohoku University, Sendai 980-8578, Japan}

\author[0000-0002-3517-1956]{Kunihito Ioka}
\affiliation{Yukawa Institute for Theoretical Physics, Kyoto University, Kyoto 606-8502, Japan}

\author[0000-0003-4299-8799]{Kazumi Kashiyama}
\affiliation{Astronomical Institute, Graduate School of Science, Tohoku University, Sendai 980-8578, Japan}

\author[0000-0001-8253-6850]{Masaomi Tanaka}
\affiliation{Astronomical Institute, Graduate School of Science, Tohoku University, Sendai 980-8578, Japan}
\affiliation{Division for the Establishment of Frontier Sciences, Organization for Advanced Studies, Tohoku University, Sendai 980-8577, Japan}

\begin{abstract}
Recent observations indicate that stripped-envelope core-collapse supernovae are often surrounded by dense circumstellar material (CSM). 
Motivated by this, we develop an analytic model to systematically study the dynamics of long gamma-ray burst (LGRB) jet propagation in various CSM environments. 
We derive a general expression for the jet head velocity ($\beta_{\rm h}$) and breakout time ($t_{\rm b}$) valid across Newtonian, relativistic, and intermediate regimes, accounting for a previously unrecognized dependence on $1 - \beta_{\rm h}$.
Our results highlight a fundamental distinction between jet propagation in massive stars, where $\beta_{\rm h}\ll 1$, and in extended CSM, where $1-\beta_{\rm h}\ll 1$. 
We establish an analytic success/failure criterion for jets and express it in terms of jet and CSM parameters, revealing a strong dependence on CSM radius. 
To quantify the relativistic nature of the jet-cocoon system, we introduce the energy-weighted proper velocity $\overline{\Gamma\beta}$. 
We identify three possible jet outcomes—(a) successful jets ($\overline{\Gamma\beta} \sim 10-100$), (b) barely failed jets ($\overline{\Gamma\beta} \sim 1$), and (c) completely failed jets ($\overline{\Gamma\beta} \sim 0.1$)—and constrain their respective jet/CSM parameter spaces. 
We show that in (b) and (c), large CSM radii can result in luminous fast blue optical transients via cocoon cooling emission.
This theoretical framework provides a basis for future observational and theoretical studies to understand the link between LGRBs, intermediate GRBs, low-luminosity LGRBs, and their environments.
\end{abstract}


\section{Introduction}
\label{sec:1}
Gamma-ray bursts (GRBs) are observed in the form of prompt emission ($\sim 1 -100$ s), making them the most energetic electromagnetic transients in the Universe (\citealt{1973ApJ...182L..85K}). 
GRBs are understood to originate from highly relativistic jets (\citealt{1997ApJ...487L...1R,2004RvMP...76.1143P}). 
In the collapsar model, long-duration GRBs (LGRBs) are explained by the core-collapse of massive stars, forming an accretion disk around a central compact object, which subsequently powers relativistic jets (\citealt{1993ApJ...405..273W,1999ApJ...524..262M}). Consequently, LGRBs are expected to be associated with the deaths of massive stars.

Follow-up observations of LGRBs have confirmed their connection to broad-lined (BL) Type Ic supernovae (SNe), such as SN 1998bw (e.g., \citealt{1998Natur.395..670G,1998Natur.395..672I,1998Natur.395..663K}) and SN 2003dh (e.g., \citealt{2003ApJ...591L..17S}), etc., providing strong observational support for the collapsar model.
These GRB-SN associations suggest that the progenitors are stripped-envelope stars, i.e., Wolf-Rayet (WR) stars. This is consistent with the collapsar scenario, as such stripped stars are compact, making them easier to penetrate by relativistic jets launched at their cores, given the expected short accretion timescales and observed emission timescales ($\sim10^2$ s; \citealt{2003MNRAS.345..575M,2011ApJ...740..100B,2011ApJ...739L..55B,2011ApJ...726..107S}). 
Numerical simulations have confirmed that jet propagation through compact progenitors is feasible under these conditions (e.g., \citealt{2000ApJ...531L.119A,2004ApJ...608..365Z,2012ApJ...750...68L,2013ApJ...777..162M,2017MNRAS.469.2361H,2018MNRAS.477.2128H,2019ApJ...871L..25P,2021MNRAS.500.3511G,2022MNRAS.517..582E,2023MNRAS.519.1941P}).

Observations have also revealed a population of low-luminosity GRBs (LLGRBs), which are significantly fainter than their high-energy counterparts (e.g., GRB 060218/SN 2006aj; GRB 100316D/SN 2010bh; \citealt{2006Natur.442.1008C,2011MNRAS.411.2792S}), 
but are significantly more common than LGRBs due to their closer proximity (\citealt{2006Natur.442.1014S,2007ApJ...662.1111L}). 
Unlike cosmological LGRBs, LLGRBs have been proposed to arise from jets that fail to break out because they are choked within dense stellar or circumstellar environments (\citealt{2015ApJ...807..172N}), leading to much weaker prompt emissions possibly due to shock breakout (e.g., \citealt{2006Natur.442.1008C,2007ApJ...667..351W,2012ApJ...747...88N}).

Consequently, the presence of an extended circumstellar medium (CSM) has been proposed as a key factor in explaining LLGRBs (\citealt{2015ApJ...807..172N,2022ApJ...925..148S,2024PASJ...76..863S}; or, alternatively to account for delayed emission in LGRBs \citealt{2024ApJ...976...55P}). 
Furthermore, the potential for bright cooling emission from their jet cocoons has been pointed out (\citealt{2013ApJ...778...67N,2015ApJ...807..172N,2022ApJ...925..148S}). 
Therefore, it has been discussed that the interaction of the relativistic jet with the CSM could produce distinct observational signatures, including fast blue optical transients (FBOTs) (\citealt{2019ApJ...872...18M,2022MNRAS.513.3810G,2024PASJ...76..863S}).
In fact, rapid follow-up observations of LGRBs / LLGRBs detected by the recent Einstein Probe (EP) mission (sensitive to softer events, in $\sim 0.5-4$ keV) show a clear association with FBOTs
(EP240414a: \citealt{2024arXiv241002315S,2025ApJ...978L..21S,2025ApJ...982L..47V,2024arXiv240919055B}; and EP250108a: \citealt{2025arXiv250408886E,2025arXiv250408889R,2025arXiv250417516S,2025arXiv250417034L}).

Observations of SNe associated with stripped-envelope progenitors (Types Ib, Ic, Ibn, and Icn), in the pre-SN or post-SN phase,
revealed the presence of CSM typically with masses $\sim 0.01-1 M_\odot$ extended over $\sim 10^{14} - 10^{16}$ cm
(see  \citealt{2007Natur.447..829P,2007ApJ...657L.105F,2008MNRAS.389..131P,2008ApJ...674L..85I,2006ApJ...653L.129B,2020A&A...643A..79S,2020MNRAS.492.2208C,2024ApJ...977..254D,2024ApJ...977....2P}; also see \citealt{2022ApJ...927...25M}).
This suggests that stripped massive stars (WR, similar stars to LGRB progenitors) undergo enhanced mass loss episodes in the last years before their explosion. 

CSM has also been observed in Type II SNe (e.g., \citealt{2006ApJ...651.1005S,2008ApJ...686..467S,2013Natur.494...65O,2014MNRAS.438.1191S,2014ApJ...797..118F,2017NatPh..13..510Y,2018NatAs...2..808F,2021ApJ...907...99S}; also, in the recent SN II 2023ixf \citealt{2023ApJ...954L..42J,2023ApJ...956L...5B,2023ApJ...956...46S}), which are more common than type Ibc SNe (\citealt{2011MNRAS.412.1441L}).
This points to the ubiquity of CSM and enhanced mass-loss CCSNe at the end of the life of massive stars in general.

The mechanism of this intense mass loss is often attributed to enhanced energy generation through nuclear burning at the late stage of stellar evolution (e.g., \citealt{2011ApJ...733...78A,2014ApJ...785...82S,2012MNRAS.423L..92Q}; \citealt{2021ApJ...915...80L}), or to binary interaction (\citealt{2017ApJ...840...90O,2018ApJ...854L..14K}).

The ubiquity of CSM around massive stars motivates the investigation of how a given CSM influences collapsar jet propagation and the resulting observational signatures. While numerical simulations could be used to explore this, the large CSM radii make such simulations numerically challenging. 
Therefore, developing an analytic framework is important.

In this work, we develop a generalized model for jet propagation in extended CSM to systematically explore how different CSM configurations influence jet dynamics, breakout conditions, and the observational diversity of GRBs and associated transients.
This paper is structured as follows. 
In Section \ref{sec:2}, we describe the physical setup. 
In Section \ref{sec:analytic}, we present our new analytic results.
In Section \ref{sec:crit}, we present the criterion for jet success or failure.
In Section \ref{sec:int jets}, we discuss the intermediate case between successful and failed jets. 
In Section \ref{sec:div}, we discuss the implications on LGRBs, intermediate GRBs, LLGRBs, and associated transients.
In Section \ref{sec:diss}, we discuss our results. 
In Section \ref{sec:con}, we present our conclusions.


\section{Model Setup}
\label{sec:2}
\subsection{Setup}
\label{sec:setup}
\begin{figure}
    \centering
    \includegraphics[width=0.99\linewidth]{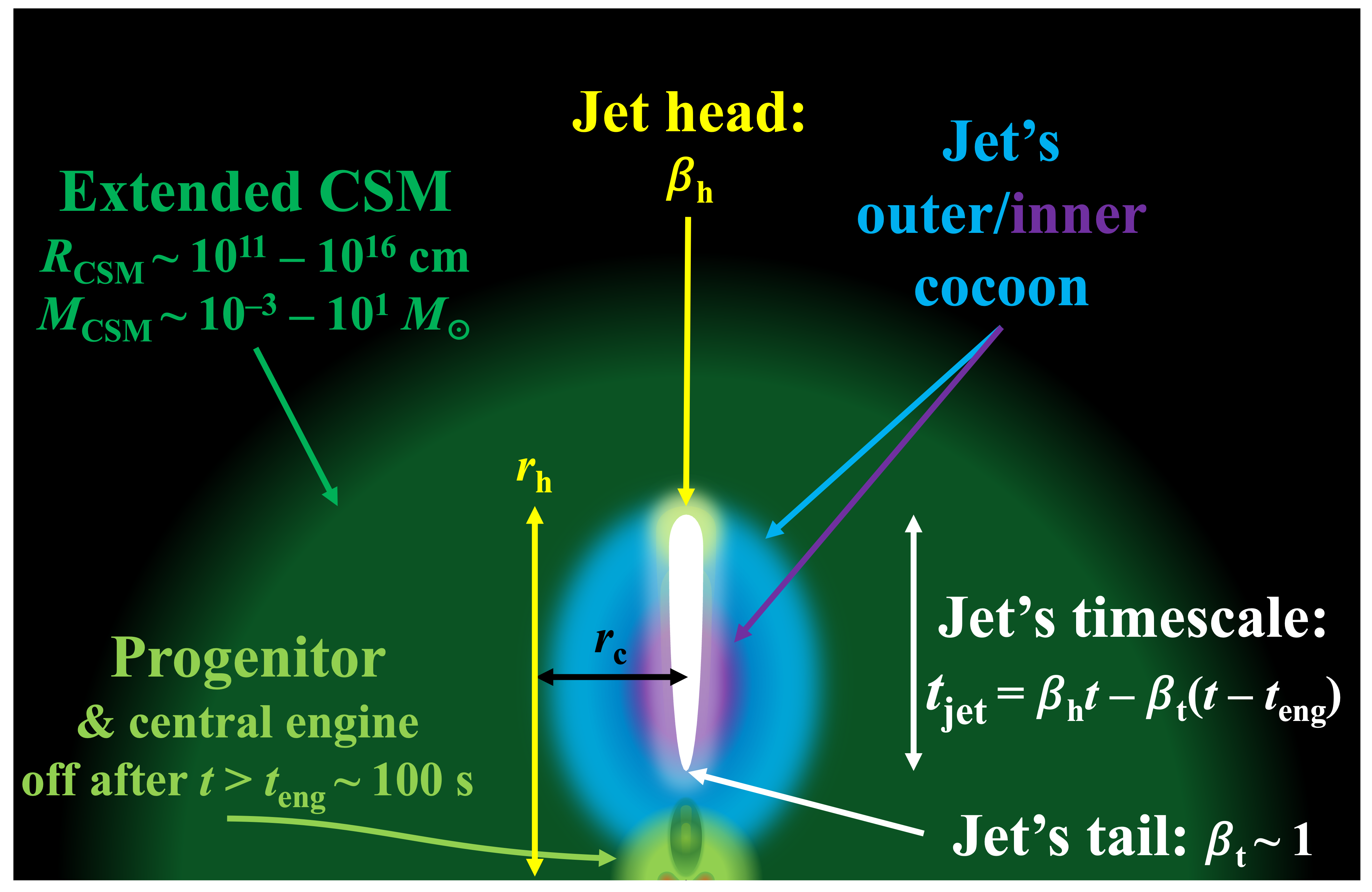} 
  \caption{Illustration of the model setup and basic concepts.
  A LGRB jet (white) is launched from the compact progenitor (light green).
  The progenitor is surrounded by an extended CSM (dark green) characterized by its total mass ($M_{\rm CSM}$) and its radius ($R_{\rm CSM}$).
  After punching through the compact progenitor, the jet propagates through the CSM, producing a jet head (yellow) and a cocoon (inner and outer, in purple and blue, respectively) which collimates the jet [see equations (\ref{eq:S1}), (\ref{eq:S2}), (\ref{eq:S3}), and (\ref{eq:S4})].
  After $t_{\rm eng}\sim 100$ s, the engine is turned off, but the jet survives inside the CSM and continues to propagate as long as the jet's tail ($\beta_{\rm t}\sim 1$, in white) does not catch up to the jet head ($\beta_{\rm h}$, in yellow).
  We define the jet's timescale which is the physical size of the jet at a given time $t$ in units of its speed $c$ [see equation (\ref{eq:tjet})].
  Once the jet's tail caches up to the jet head, the jet fails and $t_{\rm jet}=0$ [defined by equations (\ref{eq:criteria})].
  Also shown is the cocoon lateral width $r_{\rm c}$ and the jet head radius $r_{\rm h}$ [see equations (\ref{eq:betah}) and (\ref{eq:betaperp const})].
  }
  \label{fig:picture} 
\end{figure}

Figure \ref{fig:picture} illustrates the setup and key concepts referred to throughout the paper.
We consider the collapsar scenario, where the progenitor is a stripped compact ($\sim 10^{10}-10^{11}$ cm) massive WR star (\citealt{1993ApJ...405..273W,1999ApJ...524..262M}).
The progenitor is considered to be surrounded by an extended CSM, which was supposedly produced prior to the collapsar in an intense mass-loss phase.
Such CSMs have originally been discussed as a cause of jet failure and an explanation for low luminosity GRBs (\citealt{2015ApJ...807..172N}).
Here, we are interested in the dynamics of jet propagation in different CSMs and their impact on the jet propagation and emission.
Therefore, for convenience, we consider a conventional LGRB jet, while we vary the parameters of the CSM in terms of radius ($R_{\rm{CSM}}$) and mass ($M_{\rm{CSM}}$) (see Section \ref{sec:Main approximations}).

We analytically solve the collapsar-jet propagation in a given CSM, by finding key quantities, in particular the jet head velocity $\beta_{\rm h}$.
We use the analytical model in \cite{2021MNRAS.500..627H}, which was based on previous works (e.g., \citealt{2003MNRAS.345..575M}; \citealt{2011ApJ...740..100B}; \citealt{2013ApJ...777..162M}; \citealt{2018MNRAS.477.2128H}; \citealt{2020A&A...636A.105S}; \citealt{2020MNRAS.491.3192H})
with several improvements to self-consistently solve the jet propagation.

\subsection{Jump conditions}
\label{sec:Jump conditions}
The jet head dynamics in a dense ambient medium can be determined by the shock jump conditions at the jet's head (e.g., \citealt{1989ApJ...345L..21B}; \citealt{1997ApJ...479..151M}; \citealt{2003MNRAS.345..575M}):
\begin{eqnarray}
h_{\rm j} \rho_{\rm j} c^2 (\Gamma\beta)_{\rm jh}^2 + P_{\rm j} = h_{\rm a} \rho_{\rm a} c^2 (\Gamma\beta)_{\rm ha}^2 + P_{\rm a} .
\label{eq:jump}
\end{eqnarray}
Here, $h$, $\rho$, $\Gamma=(1-\beta^2)^{-1/2}$, and $P$ denote the enthalpy, mass-density, Lorentz factor (where $\beta$ is the velocity normalized by the speed of light $c$), and pressure of each fluid element, all measured in the fluid's rest frame. 
The subscripts ``${\rm j}$", ``${\rm h}$", and ``${\rm a}$" correspond to the relativistic jet, the jet head, and the cold ambient medium (CSM here), respectively.
The subscript ``${\rm jh}$" corresponds to the relative velocity between the jet and the jet head, and ``${\rm ha}$" corresponds to the relative velocity between the jet head and the ambient medium.
Since both $P_{\rm a}$ and $P_{\rm j}$ in equation (\ref{eq:jump}) are negligible, the jet head velocity can be determined as (for details, see \citealt{2018PTEP.2018d3E02I}; \citealt{2020MNRAS.491.3192H,2020A&A...636A.105S}).

\begin{eqnarray}
\beta_{\rm h}  =  \frac{\beta_{\rm j} - \beta_{\rm a}}{1 + \tilde{L}^{-1/2}} +  \beta_{\rm a} ,
\label{eq:betah 1}
\end{eqnarray}
where $\tilde{L}$ is the ratio of energy density between the jet and the ejecta,
$\tilde{L}  =  \frac{h_{\rm j} \rho_{\rm j} \Gamma_{\rm j}^2}{h_{\rm a} \rho_{\rm a} \Gamma_{\rm a}^2}$,
which can be approximated as:
\begin{eqnarray}
\tilde{L} \simeq \frac{L_{\rm j}}{\Sigma_{\rm j}(t) \rho_{\rm a} c^3 \Gamma_{\rm a}^2}   ,
\label{eq:L expression approx}
\end{eqnarray}
where $L_{\rm j}$ is the jet power (per hemisphere). 
$\Sigma_{\rm j}(t)$ is the jet cross-section which can be found based on the jet collimation regime (see Section \ref{sec:jet collimation}).

\subsection{Jet collimation}
\label{sec:jet collimation}
We follow the same treatment of \citet{2011ApJ...740..100B}.
The unshocked jet's height $\hat{z}$ can be written as a function of the jet luminosity $L_{\rm j}$ and the cocoon's pressure $P_{\rm c}$:
\begin{eqnarray}
  \hat{z} = \sqrt{\frac{L_{\rm j}}{\pi c P_{\rm c}}} + z_*.
  \label{eq:z}
\end{eqnarray}
With $r_{0}$ being the inner radius at which the jet enters the medium (CSM here), $z_* = \max[r_{0},z(P_{\rm c} = P_{\rm j0})]$ is the radius beyond which the cocoon pressure dominates over the jet pressure  (\citealt{2011ApJ...740..100B}).
Initially, $z_* \sim r_0 \ll R_{\rm CSM}$ (see Section \ref{sec:Main approximations} for an idea about $r_0$ and $R_{\rm CSM}$) (see \citealt{2021MNRAS.500..627H}).

At a certain time $t$, the jet is not collimated if the radius of the jet head, $r_{\rm h}(t)$, $<\hat{z}/2$ while it is considered collimated if $>\hat{z}/2$ (see Figure 2 in \citealt{2011ApJ...740..100B}). 
Hence, the jet head's cross-section can be found for the two modes as follows:
\begin{equation}
  \Sigma_{\rm j}(t) =
    \begin{cases}
      \pi r_{\rm h}^2(t)\theta_0^2 & \text{if $r_{\rm h}(t)<\hat{z}/2$ (uncollimated jet)} ,\\
      \pi r_{\rm h}^2(t)\theta_{\rm j}^2(t) & \text{if $r_{\rm h}(t)>\hat{z}/2$ (collimated jet)} ,
    \end{cases}       
    \label{eq:Sigma}
\end{equation}
where $\theta_0$ is the initial opening angle of the jet, and $\theta_{\rm j}(t)$ is the opening angle of the jet head at a given time $t$ (after taking into account the effect of collimation).

\subsection{Main approximations}
\label{sec:Main approximations}
Here, in our analytic modeling, we consider the following approximations: 
\begin{enumerate}
    \item  The central engine's energy output in the form of the jet with an opening angle $\theta_0$ is $E_{\rm eng}$, and the engine activity timescale is $t_{\rm eng}$. We approximate that the jet power and opening are constant during this timescale, hence the jet isotropic equivalent luminosity can be found as:
    \begin{equation}
    L_{\rm {iso,0}} =\frac{E_{\rm{eng}}}{({1-\cos\theta_0})t_{\rm{eng}}} \approx \frac{2E_{\rm{eng}}}{\theta_0^2t_{\rm{eng}}},
    \end{equation}
    and the one-hemisphere jet power is $L_{\rm j}\approx \theta_0^2L_{\rm {iso,0}}/4$ (provided that $\theta_0$ is a small angle).
    \item For convenience,  we consider one set of jet parameters corresponding to observed properties of conventional LGRB jets (considering typical isotropic equivalent luminosities and typical radiative efficiency $\eta_\gamma\sim 0.1$).
    For the central engine, we consider a total energy budget of $E_{\rm{eng}}\sim 10^{52}$ erg deposited in a timescale $t_{\rm{eng}}=100$ s in the form of a relativistic jet, which is comparable to the free-fall timescale of the WR progenitor star (\citealt{2003MNRAS.345..575M}).
    We consider a jet initial opening angle of $\theta_0=10^\circ$. 
    This is the jet opening angle after punching through the stellar envelope, where the jet is expected to open up (\citealt{2013ApJ...777..162M}).
    The corresponding jet isotropic equivalent luminosity is then $L_{\rm {iso,0}}\approx 6.67\times 10^{51} \:\text{ erg s$^{-1}$}$.
    \item The jet is assumed to be launched with a high terminal Lorentz factor $\Gamma_{\rm j}\sim 100$. Hence, we can approximate: 
    \begin{equation}
        \beta_{\rm j} \simeq 1.
    \end{equation}
    The tail velocity of the jet (see Figure \ref{fig:picture}) can also be approximated as $\beta_{\rm t}\sim 1$.
    \item We consider extended CSM. Hence, in our analytic calculations we can approximate that the outer radius of the CSM is much larger than its inner radius:
    \begin{equation}
        R_{\rm CSM} \gg r_0 ,
    \end{equation}
    where we take $r_0\sim 4\times 10^{10}$ cm, corresponds to the radius of a typical LGRB progenitor (WR star; \citealt{2006ApJ...637..914W}).
    In addition, we consider that $R_{\rm CSM}=\text{Const.}$, because the time evolution of the jet head (from launch to breakout) is much faster than the typical time evolution of the mass loss process behind the making of the CSM.
    Hence:
    \begin{equation}
        \beta_{\rm a} = 0,
        \label{eq:betaa=0}
    \end{equation}
    and $\Gamma_{\rm a} =1$ [in equations (\ref{eq:betah 1}) and (\ref{eq:L expression approx})].
    \item We consider that the CSM is produced as the result of intense mass-loss episode at the end of the progenitor's life but before the collapsar event. 
    We approximate that mass loss during this episode is constant over time, so the density profile can be expressed as:
   \begin{equation}
        \rho_{\rm a}(r,t)=\rho_0 \left[\frac{r_0}{r}\right]^n ,
        \label{eq:rhoCSM}
    \end{equation} 
    where for a constant mass loss case $n=2$ and $\rho_0=\rho_{\rm CSM}(r_0) \approx \left[\frac{M_{\rm CSM}}{4\pi r_0^2 R_{\rm CSM}}\right]$.
    Although our analytic results are specific to $n=2$ case, our modeling should also be applicable at the vicinity of this value.
    We note that the extended CSM is optically thick with respect to Thomson scattering if 
    \begin{eqnarray}\label{eq:optically_thick_condition}
    R_\mathrm{CSM} &\lesssim& \sqrt{\frac{\kappa_\mathrm{sc} M_\mathrm{CSM}}{4\pi}} \notag \\ 
    &\lesssim& 1.8\times 10^{15}\,\mathrm{cm}\,\left(\frac{\kappa_\mathrm{sc}}{0.2\,\mathrm{cm^2\,g^{-1}}}\right)^{\frac{1}{2}}\left(\frac{M_\mathrm{CSM}}{0.1\,M_\odot}\right)^{\frac{1}{2}},
    \end{eqnarray}
    where $\kappa_\mathrm{sc}$ is the scattering opacity. 
    \item The fraction of internal energy inside the cocoon is reflected in the parameter $\eta = E_{\rm c,i}/E_{\rm c}$ (with $E_{\rm c}$ as the total energy of the cocoon, and $E_{\rm c,i}$ as its internal energy). 
    Here, we take $\eta\sim1$.
    This is based on numerical simulation results for jet propagation in collapsars (see Figure 1 in \citealt{2021MNRAS.500..627H}).
    \item The pressure in the cocoon, $P_{\rm c}$, is considered constant throughout the cocoon (e.g., see \citealt{2017ApJ...834...28N}). The pressure    
    is dominated by radiation pressure, and for an adiabatic index of 4/3 it is found as:
    \begin{equation}
        P_{\rm c}= \frac{E_{\rm c,i}}{3 V_{\rm c}}, 
        \label{eq:Pc}
    \end{equation} 
    where $V_{\rm c}$ is the volume of the cocoon.
    \item The cocoon morphology in each of the two hemispheres is approximated to that of an ellipsoidal; where the ellipsoid's semi-major axis and semi-minor axis at a time $t$ are $\frac{1}{2}r_{\rm h}(t)$ and $r_{\rm c}(t)$, respectively, with $r_{\rm c}(t)$ being the cocoon's lateral width (from the jet axis) at the radius $\frac{1}{2}r_{\rm h}(t)$ (see Figure \ref{fig:picture}; also see  \citealt{2021MNRAS.500..627H,2023MNRAS.520.1111H,2023MNRAS.524.4841H}).
    The volume of the cocoon can then be found as: \begin{equation}
        V_{\rm c}=\frac{2\pi}{3}r_{\rm c}^2(t) r_{\rm h}(t),
        \label{eq:Vc}
    \end{equation}
    where typically $r_{\rm h}\gg r_{\rm c}$ (see Figure \ref{fig:picture}).
\item The velocity of the cocoon expanding laterally into the CSM can be approximated as:
\begin{equation}
    \beta_\perp =\frac{1}{c}\frac{dr_{\rm c}}{dt} \propto t^0 .
    \label{eq:betaperp const}
\end{equation}
This is based on numerical simulations showing that in the case of $n=2$ (see Figure 3 in \citealt{2021MNRAS.500..627H}).
        \item As discussed in previous works (\citealt{2013ApJ...777..162M,2018MNRAS.477.2128H,2022MNRAS.517.1640G}), the analytic description of $\tilde{L}$ in equation (\ref{eq:L expression approx}) needs to be calibrated by numerical simulations. 
    The parameter $N_{\rm s}$ is introduced to calibrate the analytic value of $\tilde L$ so that:
    \begin{equation}    
        \tilde{L}^{1/2}_{\rm c} = N_{\rm s} \tilde{L}^{1/2},
        \label{eq:Ns}
    \end{equation}
    where $\tilde{L}_{\rm c}$ is the calibrated counterpart of $\tilde{L}$.
    Here, the value of $N_{\rm s}$ is determined based on comparison with numerical simulations (\citealt{2024PASJ...76..863S}), and is assumed to be constant over time.
    Another approximation is that although $N_{\rm s}$ might depend on the value of $\tilde{L}$ (i.e., jet head velocity; see Appendix C in \citealt{2021MNRAS.500..627H}), we assume it as roughly constant over the parameter space considered.
    Here, based on comparison with numerical simulations of LGRB jet propagation in different CSM (\citealt{2022ApJ...925..148S,2024PASJ...76..863S}) we take \begin{equation}
        N_{\rm s}\sim 0.5.
        \label{eq:Ns}
    \end{equation}
\item The jet is set to enter the CSM at $t=t_0$. Since the jet propagation inside the progenitor takes a relatively short timescale ($\sim 10$~s) compared to the engine timescale $t_0\ll t_{\rm{eng}}$.
Hence, for convenience we take 
\begin{equation}
    t_0\equiv 0.
    \label{eq:t0}
\end{equation}

\end{enumerate}

As a note, our analysis considers only simple hydrodynamic jets, whereas numerical simulations suggest that magnetohydrodynamic jets are a more natural result (\citealt{1977MNRAS.179..433B}). 
3D simulations indicate that purely hydrodynamic LGRB jets propagating through dense media are susceptible to becoming unstable due to Rayleigh-Taylor instability (\citealt{2019MNRAS.490.4271M}), whereas weakly magnetized jets exhibit immunity to this instability (\citealt{2020MNRAS.498.3320G}). 
Although we do not explicitly incorporate magnetic fields in our analysis, our results may still be applicable to hydrodynamically stable jets that are expected to arise from weak magnetization.

\section{General Analytic Expression of Jet Dynamics}
\label{sec:analytic}
\subsection{Solution for the jet head velocity}
We consider the jet dynamics from the time the jet enters the CSM at $t_0$ to a given time $t$.
We defined the average jet head velocity, $\langle{\beta_{\rm h}}\rangle$, as the average jet head velocity from the jet launch $t_0$ to a given time $t$, as
\begin{equation}
\langle{\beta_{\rm h}}\rangle = \frac{r_{\rm h}(t)-r_0}{c(t-t_0)},\\
\label{eq:betah}
\end{equation}
where $t$ is laboratory time in the central engine's frame, and $r_{\rm h}(t)$ is the jet head radius at $t$ (see Figure \ref{fig:picture}).  

Following the introduction of the calibration coefficient $N_{\rm s}$, the local jet head velocity can be written as:
\begin{eqnarray}
\beta_{\rm h}  = \frac{1}{1+\tilde{L}_{\rm c}^{-1/2}} ,
\label{eq:betah 2}
\end{eqnarray}
where $\tilde{L}_{\rm c}$ can be found from equations (\ref{eq:L expression approx}) and (\ref{eq:Ns}) as a function of the jet and CSM parameters as:

\begin{eqnarray}
    \tilde{L}_{\rm c}^{1/2} \approx \left(\frac{\theta_0}{\theta_{\rm j}}\right) \left[\frac{N_{\rm s}^2 L_{\rm {iso,0}}R_{\rm CSM}}{M_{\rm CSM}c^3}\right]^\frac{1}{2},
    \label{eq:L eq2}
\end{eqnarray}
where $\theta_{\rm j}$ is the only unknown here.
The expression of $\beta_{\rm h}$ [in equation (\ref{eq:betah 2})] has often been approximated to $\beta_{\rm h}\approx \tilde{L}_{\rm c}^{\frac{1}{2}}$ (for $\beta_{\rm h}\ll 1$).
However, the parameter space (of the CSM) where such approximation is valid is extremely limited.
It is worth stressing that here we do not rely on such approximation neither do we rely on the approximation in the relativistic limit ($1-\beta_{\rm h}\ll 1$).

To find $\beta_{\rm h}$, we need to express the jet opening angle $\theta_{\rm j}$ as a function of the initial parameters, considering the cocoon collimation. 
Since the shape of the cocoon is approximated to an ellipsoidal [see Section \ref{sec:Main approximations}], $r_{\rm c}(t)$ is the lateral width of the cocoon at the radius $\frac{1}{2}r_{\rm h}(t)$. 
$r_{\rm c}(t)$ is determined using the typical lateral velocity $\beta_{\perp}$, with which the cocoon expands into the ambient medium at the radius $\frac{1}{2}r_{\rm h}(t)$ (see Figure \ref{fig:picture}). 
Considering that the medium is static [equation (\ref{eq:betaa=0})], the relationship between $P_{\rm c}$ and $\beta_\perp$ can be found as:
\begin{equation}
P_{\rm c} \approx \rho_{\rm CSM}\left({r_{\rm h}}/{2}\right) c^2 \beta_{\perp}^2 ,
\label{eq:Pc ram}
\end{equation}
where $\rho_{\rm CSM}(r_{\rm h}/2)$ is the CSM density at the radius $r_{\rm h}/2$.
Considering the evolution of the cocoon width as a function of time, energy deposition from the jet head into the cocoon, and collimation, the system of equations describing the jet-cocoon collimation can be found as follows (\citealt{2011ApJ...740..100B}):
\begin{align}
\label{eq:S1}
  r_{\rm c}(t) = & c\beta_\perp(t-t_0) , \\
\label{eq:S2}
    \beta_{\perp} =&
    \sqrt{\frac{P_{\rm c}}{{\rho}_{\rm a}(r_{\rm h}/2) c^{2}}}, \\
    \label{eq:S3}
    P_{{\rm c}} =& \eta\frac{E_{\rm c}}{3 V_{\rm c}} , \\ 
    \label{eq:S4}
    \Sigma_{\rm j}(t) =& \,\, \frac{L_{\rm j} \theta_0^2}{4 c P_{\rm c}} ,
\end{align}
where the cocoon energy $E_{\rm c} =2L_{\rm j}\left(1-\langle{\beta_{\rm h}}\rangle \right) \:(t-t_0)$,
$(t-t_0)(1-\langle{\beta_{\rm h}}\rangle) \leqslant t_{\rm eng}$ [in equation (\ref{eq:S3})], and provided that:
\begin{equation}
\theta_{\rm j} < \Gamma_{\rm h}^{-1} ,    
\end{equation}
so that the shocked CSM material is allowed to spread sideways into the cocoon (\citealt{2011ApJ...740..100B}).
This condition is satisfied as long as $\Gamma_{\rm h} \lesssim \theta_0^{-1} \lesssim 5.7$, which we find to be always the case here because $\beta_{\rm h} \lesssim 0.98$\footnote{In reality, and as shown later in equation (\ref{eq:thetaj approx}) $\theta_{\rm j}<\theta_0$. Therefore, $\Gamma_{\rm h} \lesssim \theta_0^{-1}$ is a conservative limit.}.

Using equations (\ref{eq:Vc}) and (\ref{eq:betaperp const}), for $n=2$, and $t_0=0$, the above system of equation gives:
\begin{equation}
\frac{\theta_0}{\theta_{\rm j}} = \left[ \frac{E_{\rm {eng}}R_{\rm CSM}}{2^4\eta M_{\rm CSM} c^3} \frac{1}{\beta_{\rm h}(1-\beta_{\rm h})}\right]^{-\frac{1}{4}} ,
    \label{eq:thetaj approx}
\end{equation}
which reflects the degree of jet collimation.

Combining equations (\ref{eq:betah 2}), (\ref{eq:L eq2}), and (\ref{eq:thetaj approx}) gives the following equation where $\beta_{\rm h}$ is the only unknown:
\begin{equation}
\frac{\beta_{\rm h}^3}{(1-\beta_{\rm h})^5}=C,
    \label{eq:betah final}
\end{equation}
and 
\begin{equation}
C = \frac{2^6\eta N_{\rm s}^4}{c^3}\frac{E_{\rm {eng}}}{t_{\rm eng}\theta_0^4 }\frac{R_{\rm CSM}}{M_{\rm CSM}}=\text{Const}.
    \label{eq:c}
\end{equation}

In our special case of $n=2$, $C=\text{Const.}$, hence the local jet head velocity $\beta_{\rm h}=\text{Const}$.
In other words, as long as the jet being injected behind the jet head, $\beta_{\rm h}$ is constant.
Therefore, one can write:
\begin{equation}
\beta_{\rm h} \equiv     \langle{\beta_{\rm h}}\rangle .
\end{equation}
From equation (\ref{eq:thetaj approx}) one can find that in this setup, the jet opening angle is constant over time until the jet breakout/failure. 
Another observation is that $C\propto \frac{R_{\rm CSM}}{M_{\rm CSM}}$. Hence, $R_{\rm CSM}$ and $M_{\rm CSM}$ have opposing effects on $\beta_{\rm h}$ (and $C$), and as long as the ratio $\frac{R_{\rm CSM}}{M_{\rm CSM}}=\text{Const}$, $\beta_{\rm h} = \text{Const}$ (see Section \ref{sec:betah results}). 
This is because in the case of $n=2$ we have $\rho_{\rm CSM}(r)\propto \frac{M_{\rm CSM}}{R_{\rm CSM}} r^{-n}$ [see equation (\ref{eq:rhoCSM})], therefore the ambient density (important to determine $\beta_{\rm h}$) is normalized by the ratio $\frac{M_{\rm CSM}}{R_{\rm CSM}}$.

The term $1-\beta_{\rm h}$ in equation (\ref{eq:betah final}) [which arises from equation (\ref{eq:S3})] is of significant importance here. 
Jet propagation inside massive stars is characterized by a jet head with velocities on the order of $\sim 0.1c - 0.2c$. 
Here, in the context of jet propagation in the CSM, the jet head velocity $\beta_{\rm h}$ can be significantly higher due to the relatively lower density in the CSM.
Therefore, the approximation of $\beta_{\rm h} \ll 1$ should be avoided.

Equation (\ref{eq:betah final}) cannot be solved analytically.
To solve it, we employ a numerical solution approach to solve non-linear algebraic equations. 
We employ the Brent root-finding method (\texttt{scipy.optimize.root\_scalar} with \texttt{method='brentq'} in python).
Then, we search for an approximated analytic expression by first finding the asymptotic solutions at the limits $C\ll 1$ and $C\gg 1$, and then blending them together in one function (see Appendix \ref{ap:A}).
The approximated analytic solution is found as:
\begin{equation}
\beta_{\rm h} \approx \frac{\left[C^{1/3}+\left(1 - C^{-1/5}\right) (C / 0.66)^{0.79} \right]}{1 + (C / 0.66)^{0.79}}      .
\label{eq:betah analytic}
\end{equation}

\begin{figure}
    \centering
    \includegraphics[width=0.99\linewidth]{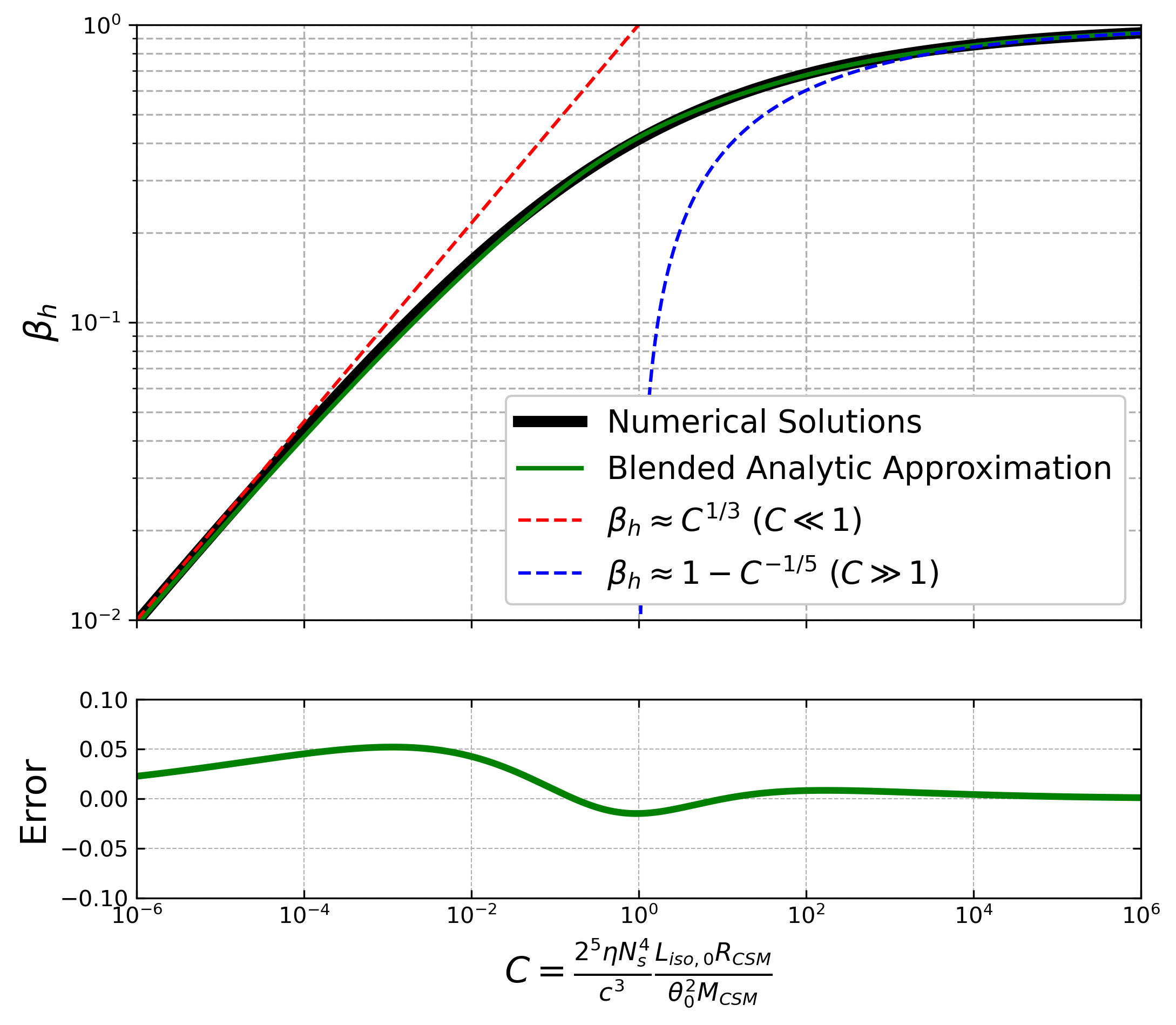} 
  \caption{The jet head velocity ($\beta_{\rm h}$, in units of $c$) as a function of the parameters of the jet and the ambient medium represented by the constant $C\propto R_{\rm CSM}/M_{\rm CSM}$.
  The numerical solution (black) [see equations (\ref{eq:betah final}) and (\ref{eq:c})], asymptotic solutions for the limits $C\ll 1$ and $C\gg 1$ (red and blue, respectively) [see equations (\ref{eq:c<<1}) and (\ref{eq:c>>1})], and the blended analytic approximation (green) [see equation (\ref{eq:betah approx})] are shown.
  Bottom panel shows the error in the blended analytic approximation.
  }
  \label{fig:blend} 
\end{figure}

In Figure \ref{fig:blend}, we show this analytic solution for $\beta_{\rm h}$ as a function of the initial conditions (of the jet and the CSM) represented by $C$, compare it to the numerical solution, and to the analytic solutions at the limits $C\ll 1$ and $C\gg 1$. 
This expression is fairly accurate ($\lesssim 4\%$) over a vast parameter space ($C\ll1$ to $C\gg 1$).
This is the first time that a general formula for describing the velocity of the jet head at both limits $\beta_{\rm h}\ll 1$ and $\beta_{\rm h}-1\ll 1$, and at all values in between.

For later use, the velocity of the cocoon lateral expansion in the CSM can also be found analytically as:
\begin{equation}
    \beta_\perp =\left[{\frac{\eta E_{\rm {eng}} R_{\rm CSM}\beta_{\rm h}(1-\beta_{\rm h})}{2^2t_{\rm eng}M_{\rm CSM}c^3}}\right]^\frac{1}{4} .
    \label{eq:betap}
\end{equation}
This description of $\beta_\perp$ is accurate as long as $\beta_\perp$ is non-relativistic [considering that it relies on equation (\ref{eq:Pc ram})], which is reasonable in the considered parameter space.

\subsection{Values of jet head velocity $\beta_{\rm h}$}
\label{sec:betah results}

\begin{figure}[htbp] 
    \centering
    \includegraphics[width=0.99\linewidth]{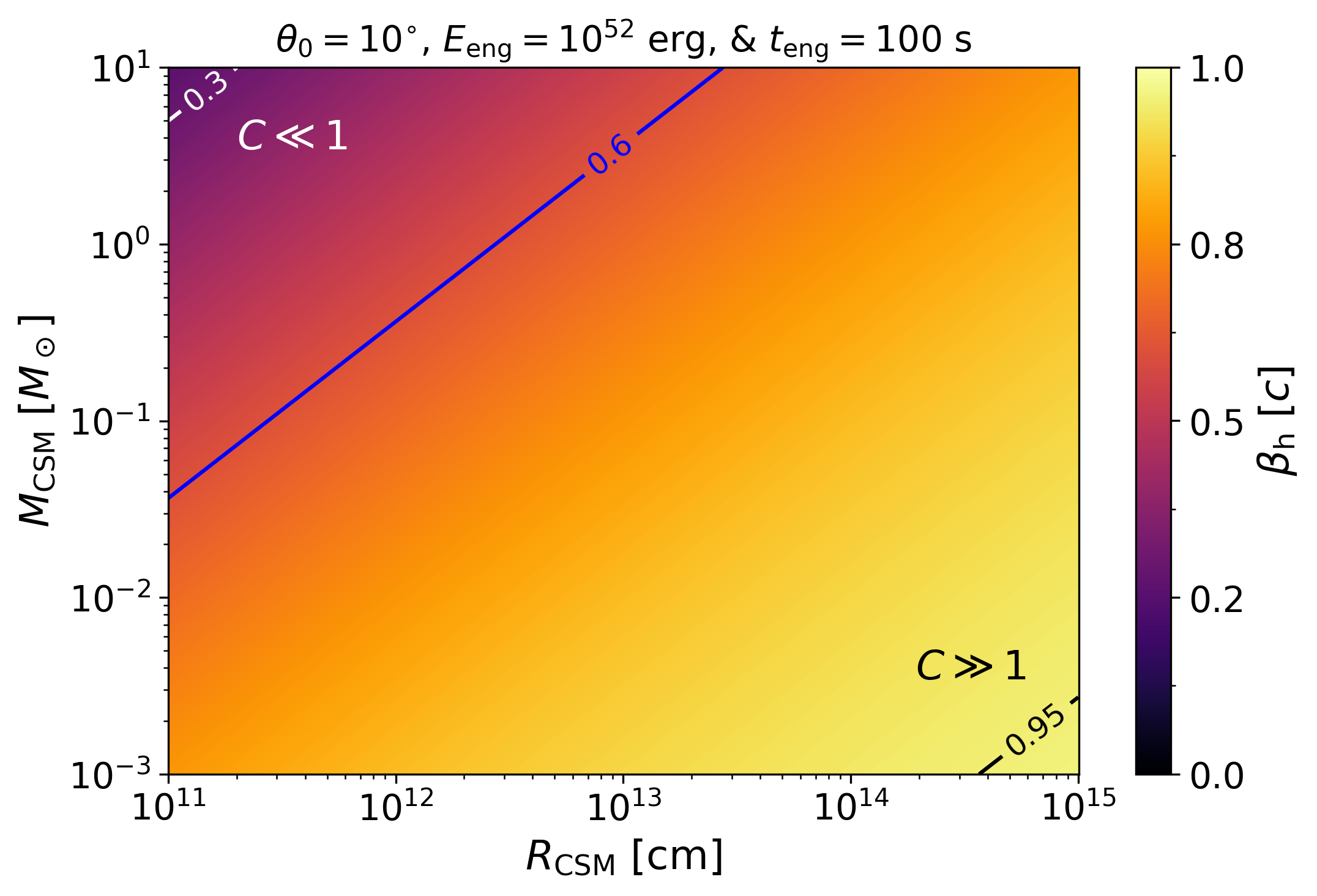} 
  \caption{Values of the jet head velocity $\beta_{\rm h}$ [see equation (\ref{eq:betah analytic})] for a wide range of CSM radii and masses, considering a conventional LGRB jet (see title). Solid lines indicate different jet head velocities $\beta_{\rm h}=0.3$ (white; upper left), $\beta_{\rm h}=0.6$ (blue), and $\beta_{\rm h}=0.95$ (black).
  The two limits where $C\ll 1$ and $C\gg 1$ are indicated [where $C\propto {R_{\rm CSM}}/{M_{\rm CSM}}$; see equation (\ref{eq:c})].
  }
  \label{fig:betah} 
\end{figure}

In Figure \ref{fig:betah}, we show the analytic solutions for $\beta_{\rm h}$ obtained using Equation~(\ref{eq:betah analytic}). 
$\beta_{\rm h}$ takes different values for various CSM with different ratios of $M_{\rm CSM}/R_{\rm CSM}$.
$\beta_{\rm h}$ varies widely, ranging from sub-relativistic values ($\beta_{\rm h} \ll 1$) to relativistic values ($\beta_{\rm h} - 1 \ll 1$).
This is because our analytic model is capable of handling both limits.

It is notable that LGRB jet propagation in CSM leads to overall relatively high jet head velocities, typically in the range $\beta_{\rm h} \sim 0.6 - 0.9$, which are significantly higher than the $\beta_{\rm h} \sim 0.1 - 0.2$ found for LGRB jet propagation inside their compact progenitors (although the possibility of a relativistic jet head has been proposed in the past; see \citealt{2001PhRvL..87q1102M}; \citealt{2001ApJ...556L..37M,2011ApJ...726..107S}).

The jet head velocity is determined by the balance of energy density between the jet and the surrounding medium. Here, the CSM masses ($M_{\rm CSM} \sim 1M_\odot$) are much lower than stellar masses ($\sim 20 M_\odot$) and are much more extended ($R_{\rm CSM} \gg r_0$), while the same jets are considered; this tips the balance in favor of the jet energy density, leading to significantly higher velocities.

Considering that previous analytical work has focused primarily on the $C \ll 1$ regime (concentrated in the upper left corner of Figure \ref{fig:betah}), our analytical formulations here allow us to explore a much wider parameter space and relate to a broader range of astrophysical phenomena.

\subsection{Breakout times}
\label{sec:tb}
Using equation (\ref{eq:betah final}) 
and $c\beta_{\rm h} (t_{\rm b}-t_0)\approx R_{\rm CSM}$, the general formula for the jet breakout time $t_{\rm b}$ can be found as:
\begin{equation}
t_{\rm b} - t_0 = \left[\frac{1}{2^6 N_{\rm s}^4\eta}\frac{R_{\rm CSM}^2 M_{\rm CSM}t_{\rm eng}\theta_0^4}{(1-\beta_{\rm h})^5 E_{\rm {eng}}}\right]^\frac{1}{3}.
\label{eq:tb1}
\end{equation}
This too, is the first time that a general formula for the breakout time has been derived -- applicable at both limits ($\beta_{\rm h}\ll 1$ and $\beta_{\rm h}-1\ll 1$), as well as all parameter space in between.

For convenience, let us take $t_0\equiv 0$ in the following.
Also, as suggested by numerical simulations $\eta\sim1$ for a collapsar jet in a static medium (see Figure 1 in \citealt{2021MNRAS.500..627H}), and we adopt $N_{\rm s}\sim 0.5$ [see equation (\ref{eq:Ns})] after comparison with simulations.
The breakout time can be found as:

\begin{equation}
\begin{split}
t_{\rm b} \approx & 430 \text{\,s}
\left(\frac{R_{\mathrm{CSM}}}{10^{13} \text{\,cm}}\right)^{\frac{2}{3}} 
\left(\frac{M_{\mathrm{CSM}}}{0.1M_\odot}\right)^{\frac{1}{3}} 
\left(\frac{\theta_{0}}{10^\circ}\right)^{\frac{2}{3}} \left(\frac{t_{\rm{eng}}}{10^2\text{\,s}}\right)^{\frac{1}{3}} \\
& \times \left(\frac{E_{\mathrm{eng}}}{10^{52} \text{\,erg}}\right)^{-\frac{1}{3}} 
\left(\frac{1-\beta_{\rm h}}{0.23}\right)^{-\frac{5}{3}} \left(\frac{N_{\rm s}}{0.5}\right)^{-\frac{4}{3}}.
\label{eq:tb apporx}
\end{split}
\end{equation}

In terms of dependence on the jet and medium parameters (CSM here), this expression is similar to the expressions found in previous work (e.g., \citealt{2011ApJ...740..100B,2015ApJ...807..172N,2018MNRAS.477.2128H,2021MNRAS.500..627H,2024ApJ...963..137H});
except for one key difference, which is the dependence on the jet head velocity term $1-\beta_{\rm h}$ discovered here for the first time.
The dependence on the term $1-\beta_{\rm h}$ has not been discovered in previous work, because, in the classical case of jet propagation in the much compact/denser environment inside massive stars, the jet head velocity is of the order of $\beta_{\rm h}\sim 0.1$ and the approximation $1-\beta_{\rm h}\sim 1$ is convenient and very reasonable [in equation (\ref{eq:S3})].
Here, because $\beta_{\rm h}-1\ll 1$, this approximation is not possible and the term $1-\beta_{\rm h}$ is highly sensitive to the value of $\beta_{\rm h}$.

The analytic values of the breakout time (as a function of the CSM mass and radius) will be discussed later in Section \ref{sec:tjets}.

\begin{figure}
    \centering
    \includegraphics[width=0.99\linewidth]{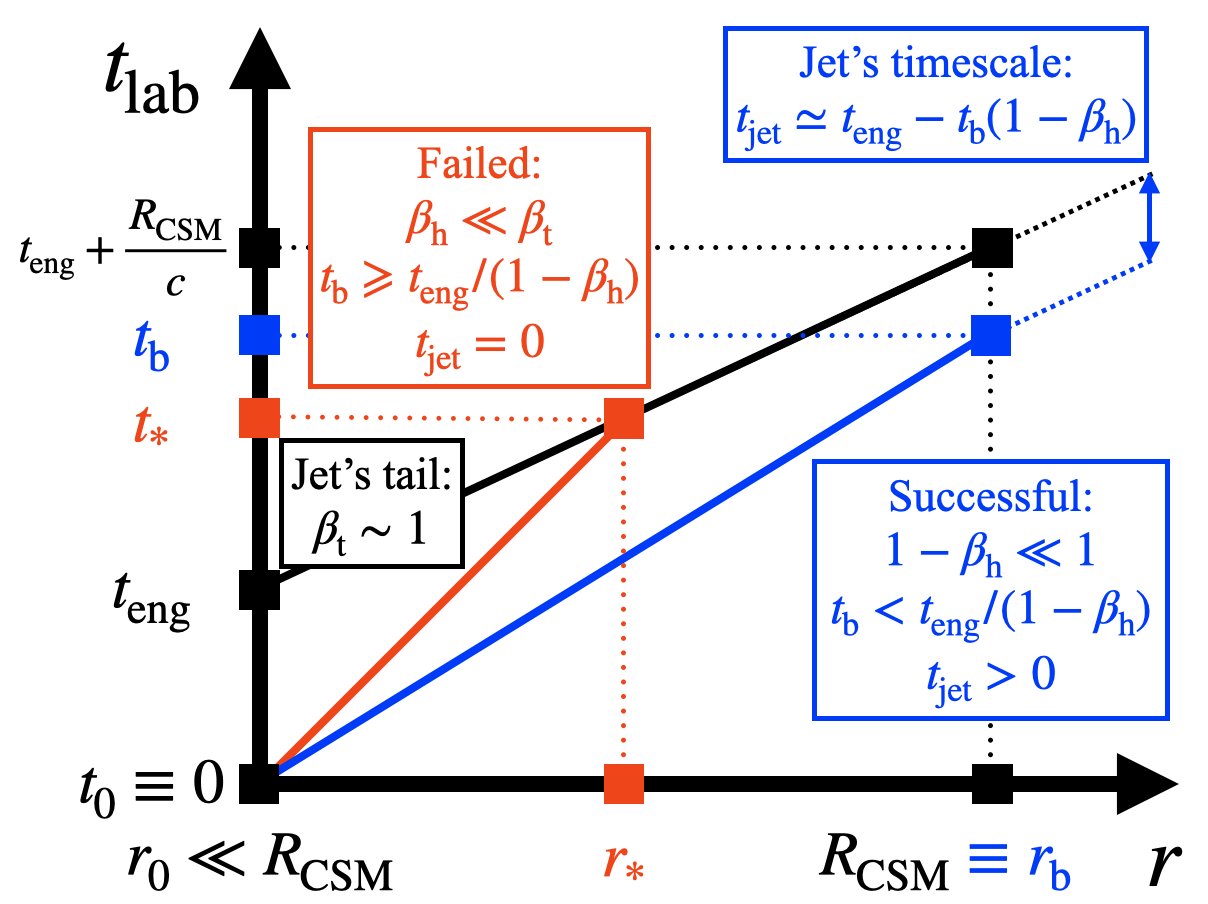} 
  \caption{Space-time diagram for jet propagation in an extended CSM. The jet is powered for a duration $t_{\rm{eng}}$. Two cases are highlighted. Successful jet (blue): the jet head velocity ($\beta_{\rm{h}}$) is high enough, so that the tail of the jet ($\beta_{\rm t}\sim 1$) does not catch up to it before the breakout. 
  As long as $t_{\rm{b}} <t_{\rm{eng}}/(1-\beta_{\rm{h}})$ part of the relativistic jet, whose timescale $t_{\rm{jet}} \simeq t_{\rm eng}-t_{\rm b}(1-\beta_{\rm h})>0$, can breakout.
  Failed jet (red; also referred as chocked): the jet head is significantly slower than the jet tail, so that the jet tail catches up with the jet head deep inside the CSM (at $t=t_*$ and $r=r_*$), and $t_{\rm jet}=0$.
  Note that due to the complexity of the gamma-ray emission process, in general $t_{\rm jet}\neq T_{90}$.
  }
  \label{fig:space-time} 
\end{figure}

\section{Criterion for Jet success/failure}
\label{sec:crit}

\subsection{Jet's timescale $t_{\rm jet}$}
\label{sec:tjet}
Figure \ref{fig:space-time} presents a space-time diagram for jet propagation in dense media.
Considering a total engine activity timescale of $t_{\rm eng}$, at a given time $t$, 
we define the ``jet's timescale" as the radial physical length of the jet (that is, the distance between the outer jet head and the inner jet tail, in the central engine rest-frame) in units of the speed of light $c$ (which is the velocity throughout the jet; see Figure \ref{fig:picture}) as:
\begin{equation}
\begin{split}
    t_{\text{jet}} \equiv \frac{\Delta r_{\rm jet} }{c\beta_{\rm j}}
    &\simeq \beta_{\rm h} t - \beta_{\rm t} (t - t_{\rm{eng}}) \\
    &\simeq t_{\rm{eng}}  -t(1 - \beta_{\rm h}) 
\end{split}
\label{eq:tjet}
\end{equation}
where the unshocked jet velocity $\beta_{\rm j}\simeq 1$, jet tail velocity $\beta_{\rm t}\simeq 1$, $t-t_{\rm{eng}}\geqslant 0$, and $t_{\rm jet} \geqslant 0$.

In the limit where the medium is too dense (or too extended) the jet's tail can catch up to the jet head inside the medium (at $t=t_*$ and $r=r_*$).
This leads to a situation where $t_{\rm jet}=0$ before the breakout, representing the condition where jet ``fails" to break out while maintaining its high terminal Lorentz factor (also called chocked or buried jet; red line in Figure \ref{fig:space-time}).
In the situation where the jet head is fast enough to breakout before the engine is turned off and the tail of the jet catches up to it, $t_{\rm jet}>0$, it is considered that the jet breaks out ``successfully" (blue line in Figure \ref{fig:space-time}); 
this jet can later contribute to the GRB prompt emission (see \citealt{2012ApJ...749..110B,2013ApJ...764..179B,2015ApJ...807..172N,2020ApJ...900..193D,2023MNRAS.519.1941P} for a similar argument) and afterglow emission due to its high Lorentz factor.

Hence, $t_{\rm jet}$ quantifies the amount of jet outflow that can emerge from the media as it was launched by the central engine, without being contaminated by baryons and without losing its highly relativistic nature ($\Gamma_{\rm j} \sim 100$).
With our assumption of constant engine power, energy in the relativistic part of the jet energy after the breakout can be found as
\begin{equation}
    E_{\rm j}(\Gamma_{\rm j}\gg 1)= \frac{t_{\rm jet}}{t_{\rm eng}}E_{\rm eng} .
    \label{eq:Ej}
\end{equation}
It should be noted that $t_{\rm jet}$ does not automatically correspond to the duration of GRB $T_{90}$, as the latter strongly depends on the radiation process and its complexity (e.g. photospheric emission: \citealt{1986ApJ...308L..43P,1986ApJ...308L..47G}; internal shock emission: \citealt{1994ApJ...430L..93R,1997ApJ...490...92K}; ICMART model: \citealt{2011ApJ...726...90Z}; etc.)

\subsection{Criterion for jet success or failure}
\label{sec:criteria}
In the following we define jet success/failure as follows:
i) ``successful" means that a fraction of the highly relativistic jet as launched by the central engine is able to successfully breakout; 
and ii) ``failed" means that the highly relativistic jet fails to breakout (see Figure \ref{fig:space-time}).

The timescale $t_{\rm jet}$ is useful for describing the fate of the jet in general terms.
The general criterion for determining the jet's fate based on $t_{\rm jet}$ can be written as:
\begin{equation}
  t_{\rm b}
    \begin{cases}
      \geqslant \frac{t_{\rm eng}}{1-\beta_{\rm h}}, & \text{$t_{\rm jet}=0$ (failed jet)} ,\\
      <\frac{t_{\rm eng}}{1-\beta_{\rm h}}, & \text{$t_{\rm jet} > 0$ (successful jet)}.
    \end{cases}       
    \label{eq:criteria}
\end{equation}
This is because once the jet head and tail intersect (at $t_* =t_{\rm eng}/(1-\beta_{\rm h})$ and $R_* \sim c\beta_{\rm h} t_*$) the scale (in lengh and/or energy) of the highly relativistic jet as injected by the central engine (with $h\Gamma_0\sim \Gamma_{\infty}\sim 100$) shrinks to $t_{\rm jet}\sim 0$.
A similar argument has been presented before (Equation 4 in \citealt{2011ApJ...726..107S}; also see \citealt{2001PhRvL..87q1102M,2001ApJ...556L..37M};  \citealt{2012ApJ...749..110B,2015ApJ...807..172N,2023MNRAS.519.1941P}).

Traditionally, the criterion for the success or failure of the jet was based on the comparison between $t_{\rm b}$ and $t_{\rm{eng}}$; that is, if $t_{\rm b}<t_{\rm{eng}}$ the jet was considered successful and if $t_{\rm b}\geqslant t_{\rm{eng}}$ the jet was considered failed (e.g., \citealt{2003MNRAS.345..575M,2011ApJ...739L..55B,2012ApJ...749..110B,2013ApJ...764..179B,2018ApJ...866L..16M}), which implicitly assumed $\beta_{\rm h}\ll 1$.
The criterion in equation (\ref{eq:criteria}) is more general, as it does not rely on any assumptions regarding $\beta_{\rm h}$.
It can be applied to the limit $\beta_{\rm h}-1\ll 1$,  to the traditional $\beta_{\rm h}\ll 1$, and intermediate cases ($0<\beta_{\rm h}<1$).

The above criterion shows that the jet head can breakout successfully despite the engine being turned off ($t_{\rm b}>t_{\rm eng}$), in particular when $1-\beta_{\rm h}\ll 1$.
This is because in such a case, the jet tail, which represents the propagation of the information that the engine has turned off, is causally disconnected with the jet head at $t\leqslant t_{\rm b}$ (considering a speed of sound $c_{\rm s}\sim c$).
Hence, in the frame of the jet head, the information that the engine has turned off does not reach the jet head at $t_{\rm b}$ and the jet head propagation continues as if the engine is still on.

\subsubsection{Criterion as a function of $M_{\rm CSM}$ \& $R_{\rm CSM}$}
\label{sec:MR crit}
This analytic criterion [as in equation (\ref{eq:criteria}) and based on the newly found solution in equation (\ref{eq:tb1})] can also be expressed analytically as a function of the CSM parameters.
Using $t_{\rm b}=t_{\rm eng}/(1-\beta_{\rm h})$, together with equations $R_{\rm CSM} \approx c\beta_{\rm h} t_{\rm b}$ and (\ref{eq:tb1}), the following alternative expression can be found:
\begin{equation}    
    M_{\rm CSM}\left(t_{\rm jet}=0\right) = D\times\left[\frac{R_{\rm CSM}^{-4}}{(1+ct_{\rm eng}/R_{\rm CSM})^2}\right],
    \label{eq:MR criteria}
\end{equation}
with $M_{\rm CSM}(t_{\rm jet}=0)\equiv M_{\rm CSM}\left(t_{\rm b }=\frac{t_{\rm eng}}{1-\beta_{\rm h}}\right)$ is the critical mass below which the jet is successful and above which the jet is failed, and
\begin{align}    
    D&=\frac{2^6c^2N_{\rm s}^4\eta t_{\rm eng}^4 E_{\rm eng}}{\theta_0^4}= \text{Const.}
\end{align}    
The dependence of this critical mass on $R_{\rm CSM}$ can then be found as:
\begin{align}
    M_{\rm CSM}\left(t_{\rm jet}=0\right) 
    \propto&
    \begin{cases}
        R_{\rm CSM}^{-4} & \text{(For $R_{\rm CSM}\gg ct_{\rm eng}$)},\\
        R_{\rm CSM}^{-2} & \text{(For $R_{\rm CSM}\ll ct_{\rm eng}$)}.\\
    \end{cases} 
\end{align}

Taking into account the typical timescales of LGRB as $t_{\rm eng}\sim 100$ s, for an extended CSM ($R_{\rm CSM}\gtrsim 10^{12.5}$ cm), we can approximate 
\begin{equation}
    R_{\rm CSM}\left(t_{\rm jet}=0\right) 
    \approx  10^{13} \text{ cm}\left(\frac{E_{\rm eng}}{10^{52}\text{ erg}}\right)^{\frac{1}{4}}  \left(\frac{M_{\rm CSM}}{10^{-1}M_{\odot}}\right)^{-\frac{1}{4}} ,
    \label{eq:MR simple}
\end{equation}
where $R_{\rm CSM}\left(t_{\rm jet}=0\right)$ is the CSM radius above/below which the jet is failed/successful, respectively.
This indicates that the critical CSM radius has the strongest effect on jet fate due to the weak dependence of other parameters on it: $R_{\rm CSM} \propto M_{\rm CSM}^{-\frac{1}{4}} E_{\rm eng}^{\frac{1}{4}}$ ($t_{\rm eng}$ also have the same effect, although in the context in LGRBs and free-fall of massive stars $t_{\rm eng}\sim 100$ s; \citealt{2003MNRAS.345..575M}).
This simple formula allows us to visualize the effect of variation in the CSM radius/mass on the fate of the jet, even beyond the considered conventional parameters (e.g., even for weak jets).

\begin{figure*}
    \centering
    \includegraphics[width=0.49\linewidth]{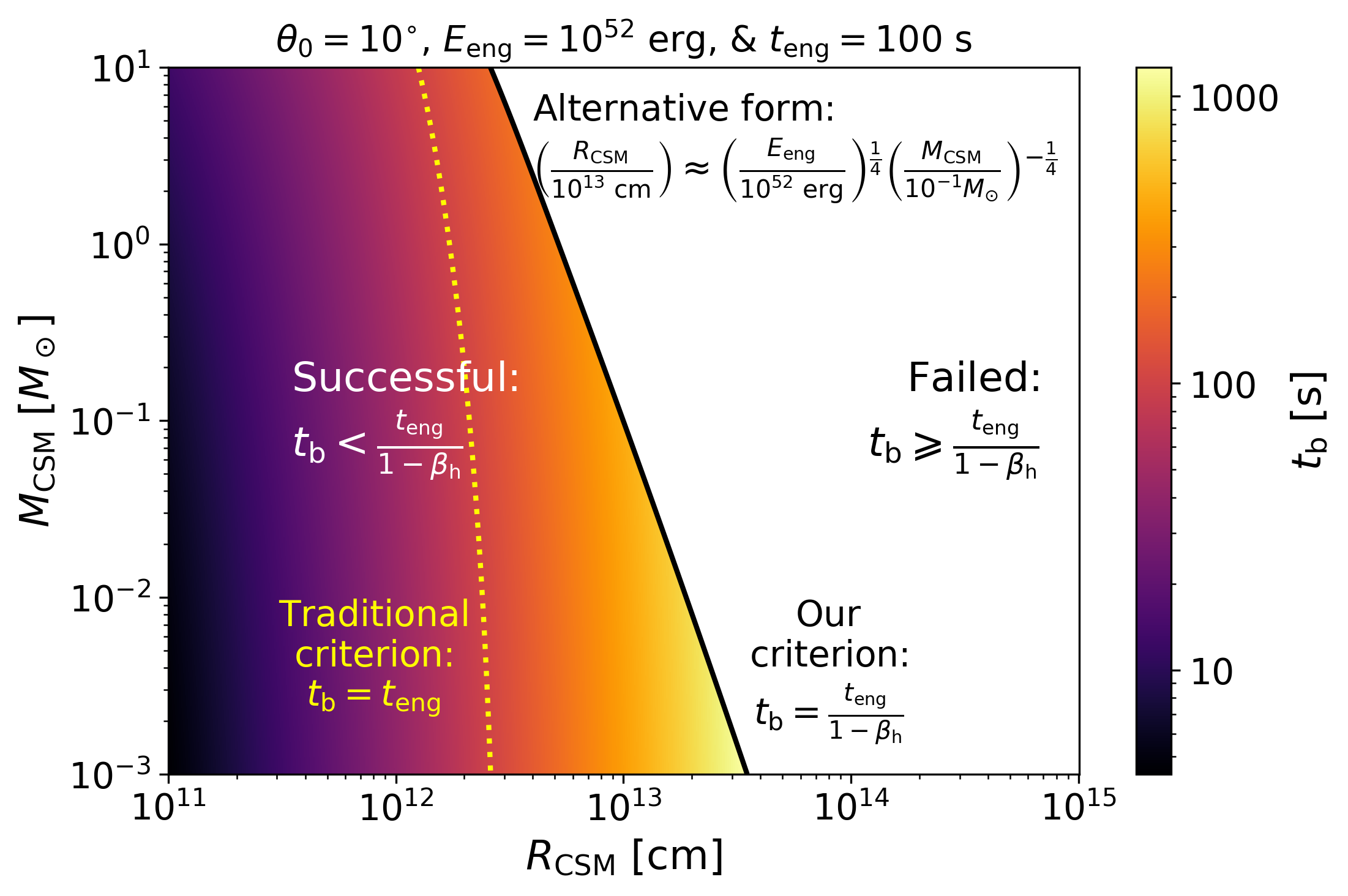} 
    \centering
    \includegraphics[width=0.49\linewidth]{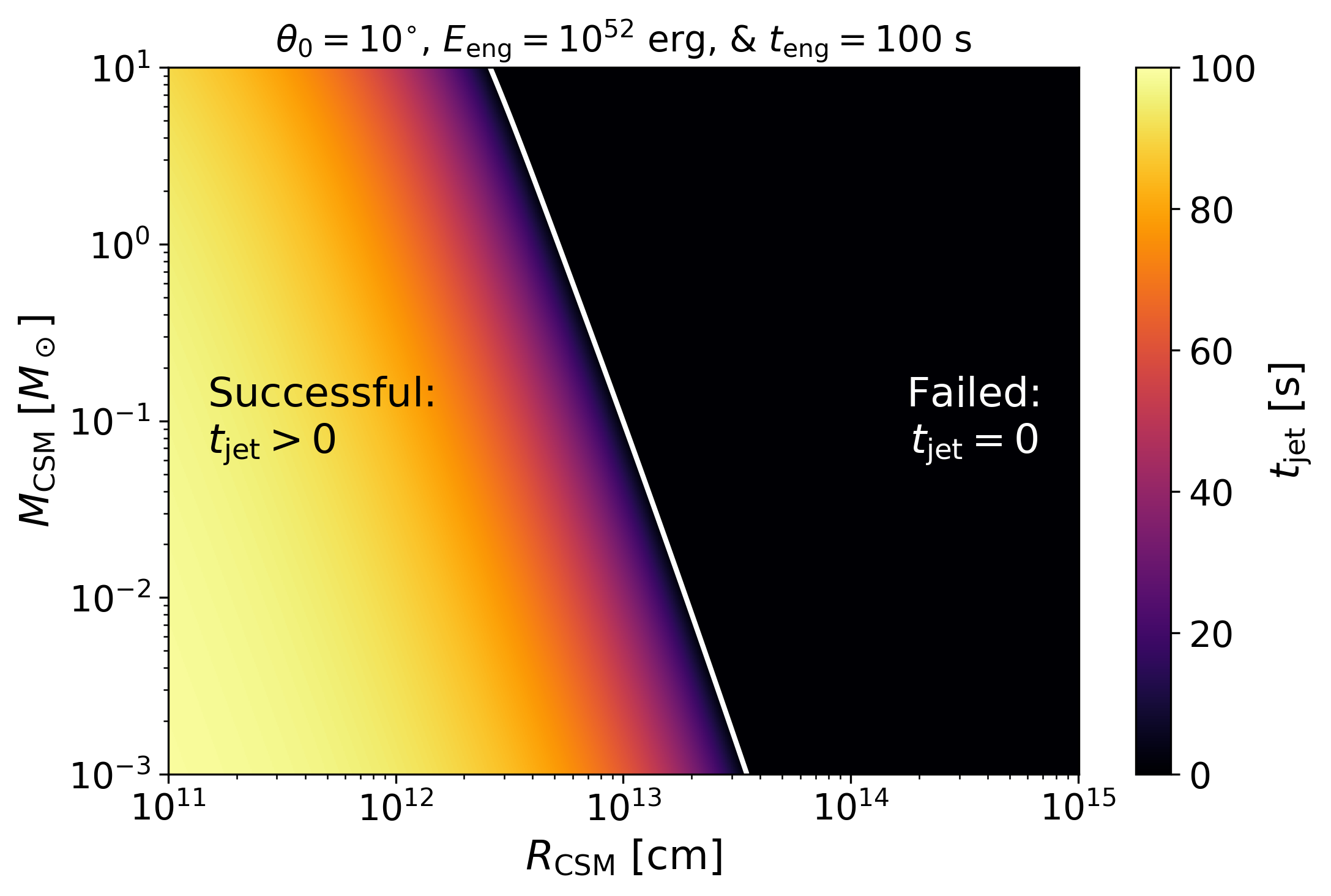} 
  \caption{Breakout times (left) and final jet timescales (right) for different CSM in terms of radius and mass [calculated using equations (\ref{eq:tb1}) and (\ref{eq:tjet}); respectively].
  The solid line (black/white) indicates the boundary between jet success and failure [equation (\ref{eq:MR criteria}) and $t_{\rm jet}=0$, respectively].
  Two cases are highlighted: successful jet (where the highly relativistic outflow can partially break out) and failed jet (where most of the jet energy is lost to the cocoon; see Figure \ref{fig:phase}) as determined by the criteria in equation (\ref{eq:criteria}) using the newly found equation (\ref{eq:tb1}).
  The yellow dotted line (left) indicate the traditional criterion used determine the jet success / failure based on comparing $t_{\rm b}$ and $t_{\rm eng}$ (e.g., \citealt{2011ApJ...739L..55B}).
  A conventional jet is considered (see title).
  }
  \label{fig:tb} 
\label{fig:tjet} 
\end{figure*}

\subsection{Values of breakout times \& jet's timescales}
\label{sec:tjets}
Figure \ref{fig:tb} (left panel) presents the breakout time for successful jets derived using equation (\ref{eq:tb1}).
The breakout times (and $t_{\rm jet}$ in the right panel) strongly depend on $R_{\rm CSM}$, which is understandable given that $R_{\rm CSM}=r_b \approx c\beta_{\rm h} t_{\rm b}$.
We show that the general criterion gives substantially different results compared to the traditional criteria [equations (\ref{eq:criteria}) and (\ref{eq:MR simple})].
For our conventional engine duration, $t_{\rm eng}=100$ s, the jet can break out successfully although the breakout times can be $\gg 100$~ s.

The right panel of Figure \ref{fig:tjet} shows the jet timescale.
The jet timescale is the longest ($t_{\rm jet}\sim t_{\rm eng}$) in the small $R_{\rm CSM}$ and small $M_{\rm CSM}$ limits.
This region reflects the limit where the physical presence of such CSM is hardly felt by the jet (conventional LGRB jet here).

For cases where the jet successfully breaks out and the CSM parameters are near the threshold [equation (\ref{eq:criteria})], the jet’s timescale can be significantly shorter than the engine lifetime, i.e., $t_{\rm jet}\ll t_{\rm eng}$. These cases may appear observationally as an apparently short GRB from a collapsing massive stripped star (see Section \ref{sec:SLGRBs}).

\section{Barely Failed Jets}
\label{sec:int jets}

\subsection{The nature of ``Barely failed" jet}
\label{sec:barely}
It is understood that the cocoon is composed of two distinct components (see Figure \ref{fig:picture}): the low mass-density high-enthalpy inner cocoon; and the high mass-density low-enthalpy outer cocoon (also, shocked jet and shocked medium, respectively; see: \citealt{2011ApJ...740..100B,2017ApJ...834...28N}).
Previous works (\citealt{2017ApJ...834...28N,2021MNRAS.500.3511G}; and particularly \citealt{2022MNRAS.517..582E,2023MNRAS.519.1941P} in more detail) show that the asymptotic energy distributions of the different components of the jet-cocoon system are comprehensible when expressed using $\frac{dE}{d\log{\Gamma\beta}}$.
In particular, the outer cocoon,
the inner cocoon
and the successful jet component can each be identified as different (approximately flat) components (see \citealt{2021MNRAS.500.3511G,2022MNRAS.517..582E,2023MNRAS.519.1941P}).

\begin{figure}[htbp] 
    \centering
    \includegraphics[width=0.99\linewidth]{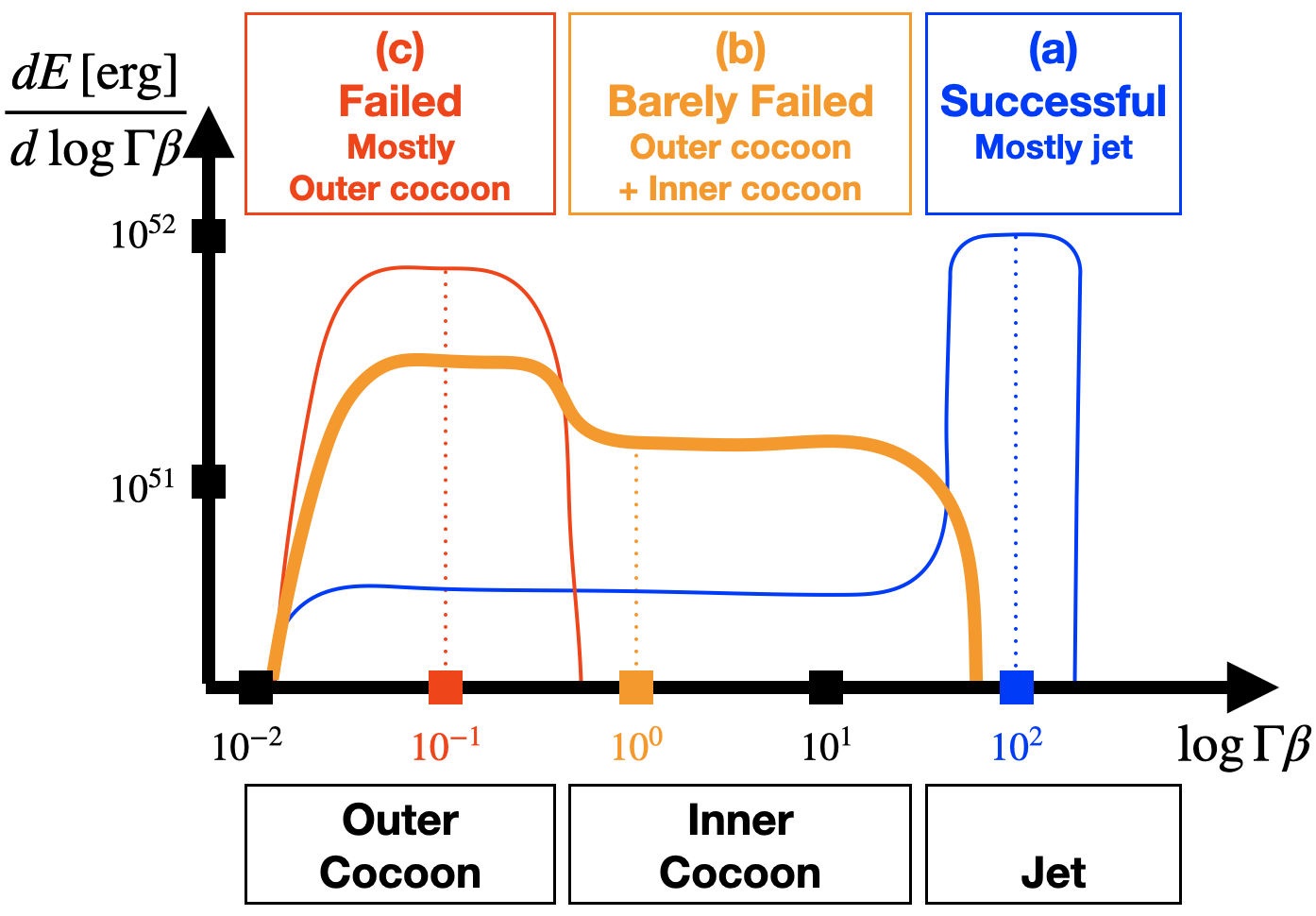} 
  \caption{Illustration energy profiles of the jet-cocoon system after the breakout. Three cases are shown: (a) successful jet (blue), (c) the failed jet (red; e.g., see \citealt{2023MNRAS.519.1941P}), and the intermediate  (b) ``barely failed" jet (orange; see \citealt{2022MNRAS.517..582E}).
  The colored vertical dotted lines indicate the typical $\Gamma\beta$ of the system for each case [found using (\ref{eq:av})].
  Lorentz factor here refers to the terminal Lorentz factor ($\Gamma\equiv \Gamma_{\infty}$).
  }
  \label{fig:jets} 
\end{figure}

In figure \ref{fig:jets}, we schematically show the expected energy distribution as a function of the proper velocity $\Gamma\beta$, for different jets.
Case (a) shows a successful jet ($t_{\rm b}\ll t_{\rm eng}/(1-\beta_{\rm h})$) which results in a strong jet that carries most of the engine energy.
Case (c) shows a completely failed jet ($t_{\rm b}\gg t_{\rm eng}/(1-\beta_{\rm h})$), where both the jet and the inner cocoon ended up being mixed with the outer cocoon, transferring all their energy to it.
Case (b) represents an intermediate scenario, where the jet fails ($t_{\rm b}\gtrsim t_{\rm eng}/(1-\beta_{\rm h})$), but moderately relativistic inner cocoon is present.
This could be the case in circumstances where the breakout takes place shortly after the jet tail catches up to the jet head, before the inner cocoon is mixed with the outer cocoon, or in cases where the jet fails in a low-density medium.
We refer to this case as ``barely failed" (after \citealt{2022MNRAS.517..582E} in the context of jet propagation inside massive stars).
Possible candidates for this case are: GRB~200826 (\citealt{2021NatAs...5..917A}), EP240414a (\citealt{2024arXiv241002315S}; \citealt{2025arXiv250316243H}), etc.

The presence of such intermediate case was observed in numerical simulations, showing that even when the head and tail of the jet intersect inside the CSM a mildly relativistic jet outflow, with terminal Lorentz factor $\Gamma_\infty \sim h\Gamma\sim 10-100$ can persist, to breakout subsequently (see Figure 6 in \citealt{2024PASJ...76..863S}; \citealt{2022ApJ...925..148S}; also in the context of stellar cores \citealt{2022MNRAS.517..582E,2023MNRAS.519.1941P}).

We attempt to find in which parameter space the intermediate case (b) occurs.
In Section \ref{sec:crit} we defined the parameter space where (a) successful jets and (c) failed jets are expected to occur.
Both (b) and (c) should fulfill $t_{\rm b}\geqslant t_{\rm eng}/(1-\beta_{\rm h})$, and $E_{\rm c} \sim E_{\rm eng}$.
As pointed out in \cite{2017ApJ...834...28N,2023MNRAS.520.1111H,2024PASJ...76..863S}, the interaction between the inner and outer cocoons is highly dependent on the degree of mixing, which is poorly understood even from numerical simulations.
An alternative is to predict where (b) is favored to occur is to quantify the relativistic degree of the whole system.
This is because of the expected strong presence of the mildly relativistic inner cocoon in (b), which helps to discriminate between (b) and (c) [(a) also] (see dotted lines in Figure \ref{fig:jets}).

\subsection{Typical $\Gamma\beta$ of the system}
We attempt to quantify the relativistic nature of the jet-cocoon system in its asymptotic phase, as a function of CSM parameters.
To characterize the relativistic nature of the jet-cocoon system, we estimate its typical proper velocity $\overline{\Gamma\beta}$, which we define as:
\begin{align}     
\overline{\Gamma\beta} = 10^{\langle{\log(\Gamma\beta)}\rangle} ,
\label{eq:av}
\end{align}
with
\begin{align}     
\langle{\log(\Gamma\beta)}\rangle
    =& \langle{\log(\Gamma\beta)}\rangle_{\rm j}\frac{E_{\rm j}}{E_{\rm eng}} + \langle{\log(\Gamma\beta)}\rangle_{\rm c} \frac{E_{\rm c}}{E_{\rm eng}},
\end{align}
where the energy-weighted logarithmic proper velocities for the jet and cocoon are given by:
\begin{align}
\langle{\log(\Gamma\beta)}\rangle_{\rm j}=&\frac{1}{E_{\rm j}}\int_{\rm jet} \log(\Gamma_{\rm j}\beta_{\rm j})\,{dE_{\rm j}}, \\
\langle{\log(\Gamma\beta)}\rangle_{\rm c}=&\frac{1}{E_{\rm c}}\int_{\rm cocoon}\log(\Gamma_{\rm c}\beta_{\rm c})\,{dE_{\rm c}}.
\end{align}

We define \( \overline{\Gamma\beta} \) in terms of the energy-weighted logarithm of the proper velocity because we want to obtain the mean value of $\Gamma\beta$ in the energy distribution \( \frac{dE}{d\log(\Gamma\beta)} \).
The energy weight enables us to visualize the energetically dominant component in the system.
Hence, this formulation provides a useful way to capture the relativistic properties of the system in a wide range spanning from the relativistic limit (e.g., (a) successful jet case with \( \Gamma\beta \sim \Gamma \sim 100 \)), 
to the non-relativistic limit (e.g., (c) failed jet case with \( \Gamma\beta \sim \beta \sim 0.1 \)),
while still being able to distinguish the intermediate case (e.g., (b) barely failed with \( \Gamma\beta \sim \beta \sim 1 \)). 

Using the analytical formulation presented in Section \ref{sec:2}, we can derive the energy of the jet and the cocoon, and their corresponding Lorentz factors.
The Lorentz factor of the jet after the breakout is expected to converge to $\Gamma_j\sim h_0\Gamma_0\sim 100$.
For the cocoon, typical velocities can also be estimated once the mass of the cocoon is estimated.
The cocoon mass can be found by integrating the mass of the shocked CSM inside the cocoon domain represented by an ellipsoid (see Section \ref{sec:Main approximations}).
The approximated expression for the cocoon mass at the breakout time has been derived in \cite{2023MNRAS.520.1111H} as:
\begin{equation}
\frac{M_c}{M_{\rm CSM}}\approx 2 \left[\ln\left(\frac{R_{\rm CSM}}{r_0}\right)-1\right] \frac{V_{\rm c}}{V_{\rm CSM}} ,
\label{eq:Mc final}
\end{equation}
where $V_{\rm c}$ can be found from equation (\ref{eq:Vc}) [and provided that $r_{\rm h}\gg r_{\rm c}$ which is generally the case here].
The typical Lorentz factor of the cocoon can then be found using:
\begin{equation}
    \Gamma_{\rm c} = 1+ \frac{E_{\rm c}}{M_{\rm c}c^2},
    \label{eq:Gammac}
\end{equation}
and the velocity can be found as $\beta_{\rm c} = \sqrt{1-\Gamma_{\rm c}^{-2}}$. 
$\Gamma_{\rm c}$ here is an asymptotic value ($\Gamma_{\rm c}\equiv \Gamma_{\rm c,\infty}$),
which reflects the relativistic nature of the combined inner and outer cocoons.
It should be noted that full mixing of the inner and outer cocoons is assumed; therefore, $\Gamma_{\rm c}$ can be considered a conservative value.

\section{Diversity of Transients}
\label{sec:div}

\begin{figure*}
    \centering
    \includegraphics[width=0.99\linewidth]{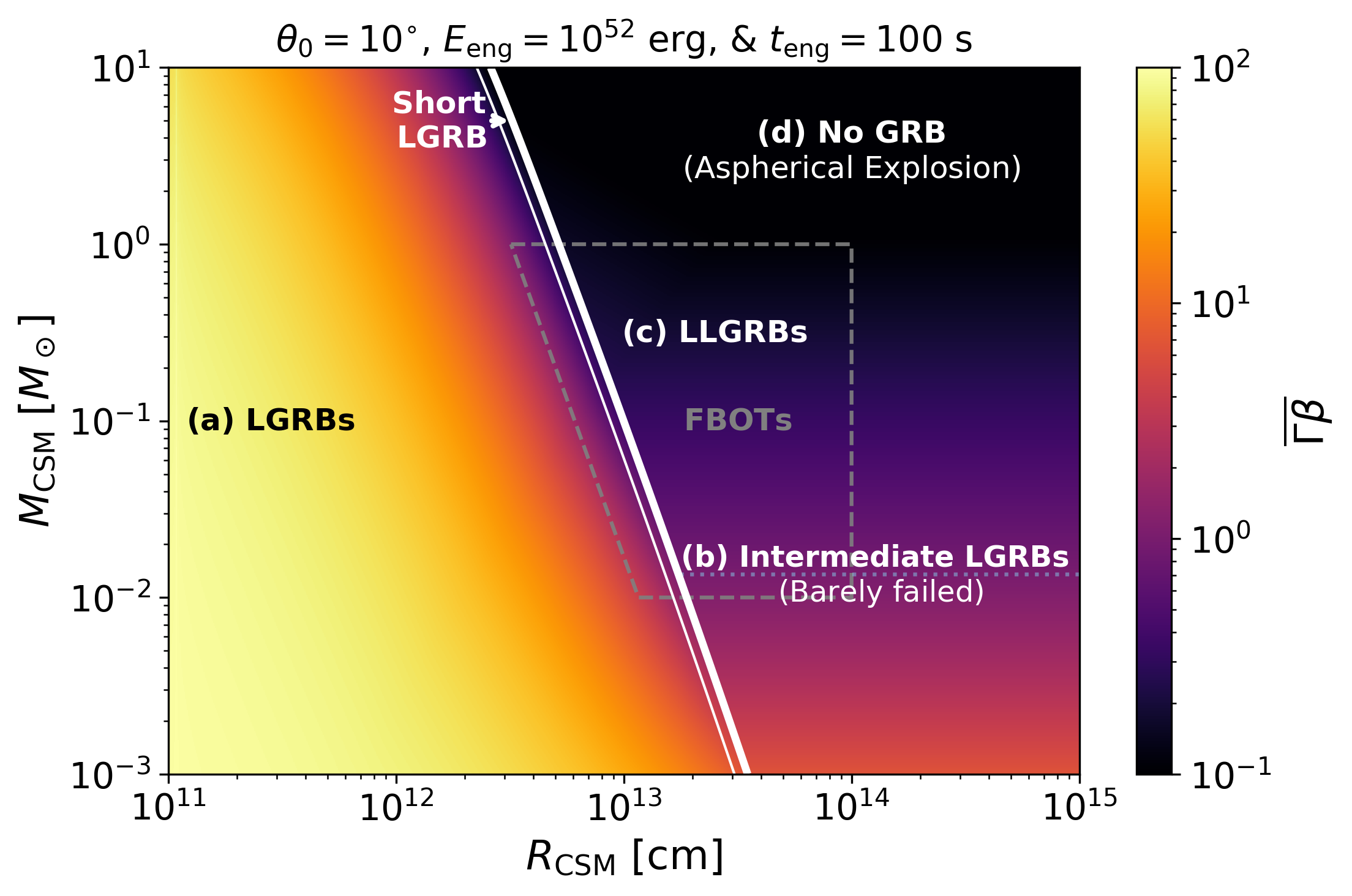} 
  \caption{
Relativistic nature of the jet outflow after breakout and the associated jet-driven transients in the CSM parameter space.
The map shows the typical proper velocity of the system [using equation (\ref{eq:av})].
Four cases are highlighted:  
(a) LGRBs ($\overline{\Gamma\beta}\sim 10 -100$): where a highly relativistic successful jet (left to the thick white line) carries the majority of the engine energy;
(c) LLGRB ($\overline{\Gamma\beta}\sim 0.1 -1$): 
where a jet fails deep inside the CSM, and a LLGRB-like shock breakout emission is expected;  
(b) barely failed jet ($\overline{\Gamma\beta}\sim 1$): an intermediate case between (a) and (c) where the jet fails but a mildly relativistic outflow ($\Gamma_{\rm j} \sim 1 - 10$) is expected to remain;
and (d) No GRB region ($\overline{\Gamma\beta}\lesssim 0.1$): as the jet fails entirely, and the cocoon is expected to produce an aspherical explosion.
The horizontal dotted light blue line indicate the region with $\overline{\Gamma\beta}= 1$.
The region where luminous thermal FBOT powered by the cocoon cooling emission, thanks to the large breakout radius and short diffusion timescales (within the dashed line) is indicated (e.g., \citealt{2019ApJ...872...18M,2024PASJ...76..863S}; and recently \citealt{2024arXiv241002315S}; \citealt{2025arXiv250417034L}).  
Luminous afterglow emission (from radio to X-ray) is expected in case (a), and to a lesser extent in case (b), and can produce non-thermal FBOT-like signatures, for both on-axis and off-axis viewing angles.
The white thick/thin solid lines are for $t_{\rm jet}=0$, and  $t_{\rm jet}=t_{\rm eng}/10$; and the region in between indicated by the white arrow is where short LGRBs could occur (e.g., GRB~200826; see \citealt{2021NatAs...5..917A}).
A conventional LGRB jet is assumed (see title) with $N_{\rm s}= 0.5$.  
An on-axis observer is considered, and viewing angle effects are not taken into account.
}
  \label{fig:phase} 
\end{figure*}

In Figure \ref{fig:phase} we present the typical asymptotic proper velocity of the jet-cocoon system [$\overline{\Gamma\beta}$, see equation (\ref{eq:av})] for a conventional LGRB jet in a wide variety of CSMs.

\subsection{LGRBs}
\label{sec:LGRBs}
The parameter space for (a) LGRBs ($\overline{\Gamma\beta}\sim 10-100$) corresponds to that of successful jets, where the highly relativistic jet dominates energetically.
This parameter space is defined by the criterion $t_{\rm b}<t_{\rm eng}/(1-\beta_{\rm h})$.
As labeled, this parameter space corresponds to that of conventional LGRBs, in terms of potentially luminous prompt emission ($E_{\rm iso,\gamma}\sim 10^{50}-10^{55}$ erg) followed by luminous afterglow emission.
Such dominant jets make these LGRBs potentially observable even at high redshift ($z \gtrsim 1$).
Using our criterion for jet success [equation (\ref{eq:MR simple})], we can constrain the CSM of these LGRBs as
\begin{equation}
    R_{\rm CSM}^{\rm LGRBs} \lesssim 10^{13}\text{\,cm}\left(\frac{E_{\rm eng}}{10^{52}\text{ erg}}\right)^\frac{1}{4}\left(\frac{M_{\rm CSM}}{10^{-1}M_{\odot}}\right)^{-\frac{1}{4}} .
    \label{eq:LGRB crit}
\end{equation}
This reveals the extreme CSM-less environment of luminous LGRBs, in striking contrast to CSM observed in SNe Ibc (and other CCSNe, see Section \ref{sec:1}).

\subsection{``Short" LGRBs}
\label{sec:SLGRBs}
For cases where the jet successfully breaks out and the CSM parameters are near the boundary of the parameter space [equation (\ref{eq:LGRB crit})], $t_{\rm jet}\ll t_{\rm eng}$. 
We show the corresponding parameter space (between the thin and solid lines indicated by the arrow in Figure \ref{fig:phase}; for $t_{\rm jet}= t_{\rm eng}/10$).
In this case, and in particular when $t_\mathrm{jet} < 2\,\mathrm{s}$, the duration of the prompt emission ($T_{90}$) can be also shorter than $2\,\mathrm{s}$ and such an event may be classified as a short GRB (\citealt{2011ApJ...739L..55B,2012ApJ...749..110B,2013ApJ...764..179B,2013RSPTA.37120273P}).
Assuming that the prompt emission occurs at $r \geqslant R_\mathrm{CSM}$ through the same mechanism as LGRBs, 
other emission properties, except for the duration, e.g., the emission spectrum, will be consistent with those of LGRBs. 
One possible example is GRB~200826 (\citealt{2021NatAs...5..917A}).

In these cases, since a significant fraction of the jet energy is converted into the internal and kinetic energy of the cocoon,
a bright shock breakout emission and an FBOT-like counterpart may also accompany the event, in particular for large $R_\mathrm{CSM}$ (see Section \ref{sec:fbot}).

\subsection{No GRB \& Aspherical Explosion}
For a sufficiently dense CSM as $M_\mathrm{CSM} \gtrsim 1\,M_\odot$,
the jet is completely failed and the system is expected to transit to the non-relativistic regime, with no possibility for a LGRB/LLGRB-like prompt emission (d).
Still, we expect an aspherical explosion powered by the non-relativistic cocoon, with an explosion energy $E_{\rm c}\sim E_{\rm eng}$ and a typical velocity $\beta_{\rm c}\sim \sqrt{2E_{\rm eng}/(M_{\rm CSM}c^2)}\sim 0.1 - 0.01$.
We expect the degree of asphericity to decrease as the CSM radius increases (from left to right in Figure \ref{fig:phase}), with the explosion becoming near-spherical in the limit where $R_{\rm CSM}$ (and $M_{\rm CSM}$) is very large.
This jet-driven explosion could be observed in the form of a BL-type SN (e.g., \citealt{2015ApJ...805..164N,2019ApJ...871L..25P,2019MNRAS.489.2844I,2022MNRAS.517..582E,2023MNRAS.519.1941P}) or in the form of an aspherical explosion (see also Section \ref{sec:weak}).

\subsection{LLGRBs}
\label{sec:LLGRBs}
In the failed cases with an intermediate-mass CSM of $\text{a few }\times10^{-2}M_{\odot} \lesssim M_{\rm CSM} \lesssim M_{\odot}$, the outer cocoon breaks out of the CSM with a trans-relativistic velocity, $\overline{\Gamma\beta}\lesssim 1$. 
The shock breakout produces soft to hard X-ray photons, with a duration determined by the shock crossing timescale at the outer edge of the CSM~\citep{2012ApJ...747...88N}.
In particular, for $R_{\rm CSM} \sim 10^{13\mbox{-}14}\,\mathrm{cm}$, the emission properties are consistent with those of the observed LLGRBs, such as GRB~060218, which has $L_\mathrm{iso, peak} \sim 10^{48}\,\mathrm{erg\,s}^{-1}$ and $T_{90} \sim 10^3\mathrm{\,s}$ (\citealt{2015ApJ...807..172N}; \citealt{2016MNRAS.460.1680I,2024arXiv241206736I}).
The lack of a strong relativistic component in this parameter space is expected to produce an afterglow emission that is much dimmer than that of LGRBs (as in GRB 060218; \citealt{2006Natur.442.1008C,2006ApJ...638..930S,2006Natur.442.1011P,2006Natur.442.1014S}).

If an intermediate-mass CSM with $\text{a few }\times10^{-2}M_{\odot} \lesssim M_{\rm CSM} \lesssim M_{\odot}$ has a more extended CSM with $R_\mathrm{CSM} \gtrsim 10^{14}\,\mathrm{cm}$ and a conventional GRB jet is failed inside it, the associated quasi-thermal shock breakout emission becomes brighter and longer than the observed LLGRBs. 
Such a transient has not been confirmed so far. 
On the other hand, such extended CSM has been inferred from observations of interaction-powered stripped-envelope SNe, or type Ibn/Icn SNe (see \citealt{2007Natur.447..829P,2007ApJ...657L.105F,2008MNRAS.389..131P,2008ApJ...674L..85I,2006ApJ...653L.129B,2020A&A...643A..79S,2020MNRAS.492.2208C,2024ApJ...977..254D,2024ApJ...977....2P}; also see \citealt{2022ApJ...927...25M}).
This implies that these observed events are from a distinctly different population, with no LGRB-like jets produced.   

\subsection{Barely failed GRBs}
\label{sec:Barely failed GRBs}
We identify the parameter space for the intermediate case between (a) and (c), in the form of (b) barely failed jets (see Figure 6 in \citealt{2022MNRAS.517..582E}).
As explained in Section \ref{sec:barely}, in this case, although the jet is failed, the cocoon is expected to be mildly relativistic.
An alternative definition to this intermediate case is that it has a potentially energetic inner cocoon component.
We consider the typical proper velocity of this case at $\overline{\Gamma\beta}\sim 1$ as the midpoint between (a) and (c) (see Figure \ref{fig:jets}) although this boundary is not robust and would depend on the degree of mixing (and on the value of $\Gamma_{\rm j}$).
From Figure \ref{fig:phase}, we expect this intermediate case (b) to be favored by less massive CSM (compared to (c)) with 
\begin{equation}
    \begin{cases}
        R_{\rm CSM}^{\rm (b)}\gtrsim R_{\rm CSM}^{\rm LGRBs},\\  
        M_{\rm CSM}^{\rm (b)}\lesssim 0.1M_\odot(E_{\rm eng}/10^{52}\text{ erg}).
    \end{cases}
    \label{eq:b cases}
\end{equation}

In addition, (b) might also occur in the boundary region between jet success and failure (solid white line in Figure \ref{fig:phase}) in the case where the jet is marginally failed but not well mixed with the outer cocoon (\citealt{2019MNRAS.490.4271M,2020MNRAS.498.3320G}).
The observational smoking gun of (b) would be: intermediate prompt emission luminosity (weaker than LGRBs but brighter than LLGRBs) and jet-like but intermediate afterglow emission.
A possible candidate for (b) is the recent EP event EP240414a (\citealt{2024arXiv241002315S,2025ApJ...978L..21S,2025ApJ...982L..47V,2024arXiv240919055B}), which is addressed in \cite{2025arXiv250316243H}.

While non-relativistic shock breakout emission from the outer cocoon can occur simultaneously, it is expected to be subdominant compared to the prompt emission from the mildly relativistic jet (and its early afterglow).
In the barely failed scenario, the mildly relativistic inner cocoon component breaks out of the CSM first, carving a path through which the bulk of the outer cocoon subsequently expands. 
As a result, the effective surface area of the outer cocoon's breakout could be reduced, which may diminish the strength of the associated shock breakout emission. 
This sequence of events may suppress its detectability.

\subsection{FBOT emission}
\label{sec:fbot}
Optical surveys have led to the discovery of rapidly evolving transients, typically characterized by durations of $\gtrsim 10$ days, luminosities of $\sim 10^{42}$--$10^{44}$ erg s$^{-1}$, and temperatures of $\sim 10{,}000$--$30{,}000$ K.  
Their event rates have been estimated to be $\sim 4000$--$7000$ Gpc$^{-3}$ yr$^{-1}$ (e.g., \citealt{2014ApJ...794...23D,2018MNRAS.481..894P,2024ApJ...977...18T}).  
More recently, a subclass of much faster transients ($\lesssim 10$ days), referred to as FBOTs, has been identified, with estimated event rates of $\sim 100$ Gpc$^{-3}$ yr$^{-1}$ (\citealt{2023ApJ...949..120H}).  
In the following, we compare the properties of the jet-driven transients discussed here with these observational populations.

In scenarios (c) and (b), the cocoon carries most of the energy of the central engine [also possibly at the boundary with (a)].  
As a result, cooling emission from the cocoon (thermal emission in the UV/optical/IR bands) is expected to be luminous, especially for large CSM radii (which reduce adiabatic losses; see dashed line in Figure~\ref{fig:phase}).  
The timescale of this cooling emission is set by the cocoon mass as  
$t_{\rm diff} \sim \sqrt{\frac{\kappa M_{\rm c}}{c^2\beta}} \propto M_{\rm c}^{1/2}$,  
where we assume an opacity $\kappa \sim 0.1$ cm$^2$ g$^{-1}$ (due to Thomson scattering), and asymptotically $M_{\rm c} \sim M_{\rm CSM}$.  
For $M_{\rm CSM} \sim 10^{-2} - 1\, M_\odot$, the corresponding diffusion timescales are fast; ranging from hours to days.
Assuming jet energies of $E_{\rm eng} \sim 10^{51}$--$10^{52}$ erg, the bolometric luminosity at $\sim 1$ day is estimated as  
$\sim \frac{t_{\rm b} E_{\rm eng}}{t_{\rm diff}^2} \sim 10^{44}$--$10^{45}$ erg s$^{-1}$ \citep{2025arXiv250316243H}.  
For a photospheric velocity of $\sim \sqrt{2E_{\rm eng}/M_{\rm CSM}}$, the Stefan–Boltzmann law gives an effective temperature of $\sim 10{,}000$–$20{,}000$ K at $\sim 1$ day, indicating a very blue color.

Comparing these features with observed transients suggests that the cocoon cooling emission discussed here is inconsistent with the rapidly evolving transients that have a timescale of $\gtrsim 10$ days (\citealt{2014ApJ...794...23D,2018MNRAS.481..894P}), due to the shorter duration (assuming $M_{\rm CSM} \sim 10^{-2}$--$1\, M_\odot$), although the color and peak luminosity are broadly consistent.  
Reproducing the longer timescales would require either a much more massive CSM ($\gtrsim 1\, M_\odot$) or a significantly more extended progenitor.  
More importantly, jet-driven explosions are intrinsically rare ($\sim$ a few $\times 100$ Gpc$^{-3}$ yr$^{-1}$; \citealt{2006Natur.442.1011P,2006Natur.442.1014S,2006ApJ...645L.113C,2007ApJ...662.1111L,2007ApJ...657L..73G}), 
and their event rates are not expected to exceed those of BL Type Ic SNe with which they are strongly associated ($\sim 5000$ Gpc$^{-3}$ yr$^{-1}$;  
\citealt{2011MNRAS.412.1441L,2017PASP..129e4201S,2020ApJ...904...35P,2023ApJ...953..179C}).

In conclusion, within the broader landscape of rapidly evolving blue optical transients, cocoon cooling emission from LGRB progenitors with extended CSM may represent the bright, blue, and fast end of the population, and could account for a significant fraction of the fastest FBOT events.
This scenario has already been proposed in the literature, particularly for the early blue peak in GRB~060218 (\citealt{2015ApJ...807..172N}; see also  
\citealt{2019ApJ...872...18M}, \citealt{2022ApJ...925..148S}, \citealt{2024PASJ...76..863S}).  
For more discussion in the context of EP240414a, see \cite{2025arXiv250316243H}.

Afterglow emission from the successful jet in (a), viewed either on or off-axis, or the mildly relativistic jet-cocoon in (b), could be very bright in optical bands and could also produce an FBOT-like emission (e.g., \citealt{2024arXiv241002315S,2025ApJ...978L..21S,2025ApJ...982L..47V,2024arXiv240919055B,2025arXiv250324266Z,2025arXiv250316243H} for the case of EP240414a; and  \citealt{2025arXiv250314588B,2025arXiv250505444G,2025arXiv250507665X,2025arXiv250508781Y} for the case of EP241021a). 
In this case emission is non-thermal, from X-ray to radio bands.
Therefore, radio/X-ray observations, together with optical observations, offer an opportunity to discriminate cooling emission (as in EP250108a; \citealt{2025arXiv250417516S}) from afterglow emission; such observations could help pin down the nature of the jet-cocoon (failed, or successful) and infer its parameters.
Hence, one could understand the reason for jet failure (e.g., in LLGRBs), as either due to intrinsically weak jet or due to thick/massive CSM (see Section \ref{sec:weak}).

\section{Discussion}
\label{sec:diss}

The observed event rates suggest that LLGRBs are $\sim 10^2$ times more abundant than LGRBs 
(see: \citealt{2005MNRAS.360L..77C}; \citealt{2006Natur.442.1011P}; \citealt{2006Natur.442.1014S}; 
\citealt{2006ApJ...645L.113C}; \citealt{2007ApJ...662.1111L}; \citealt{2007ApJ...657L..73G}; without accounting for possible differences in the beaming factors).
LLGRBs are thought to originate from failed jets, with one plausible scenario being that jet failure is caused by extended CSM \citep{2015ApJ...807..172N}. 
If jet properties are universal, this suggests that LGRB progenitors without extended/massive CSM might be rarer than those with CSM. 
However, this remains unclear due to the limited sample of stripped-envelope SNe with dense CSM, partly due to observational biases. 
One possibility is that LGRB-favored conditions (such as low metallicity; \citealt{2009ApJ...691..182S}; or high angular momentum \citealt{1999ApJ...524..262M}), correlate with progenitors lacking extended CSM. 
Alternatively, LLGRBs might result from intrinsically weak jets, as discussed in Section \ref{sec:weak}.

\subsection{Weak jet case}
\label{sec:weak}

\begin{figure}
    \centering
    \includegraphics[width=0.99\linewidth]{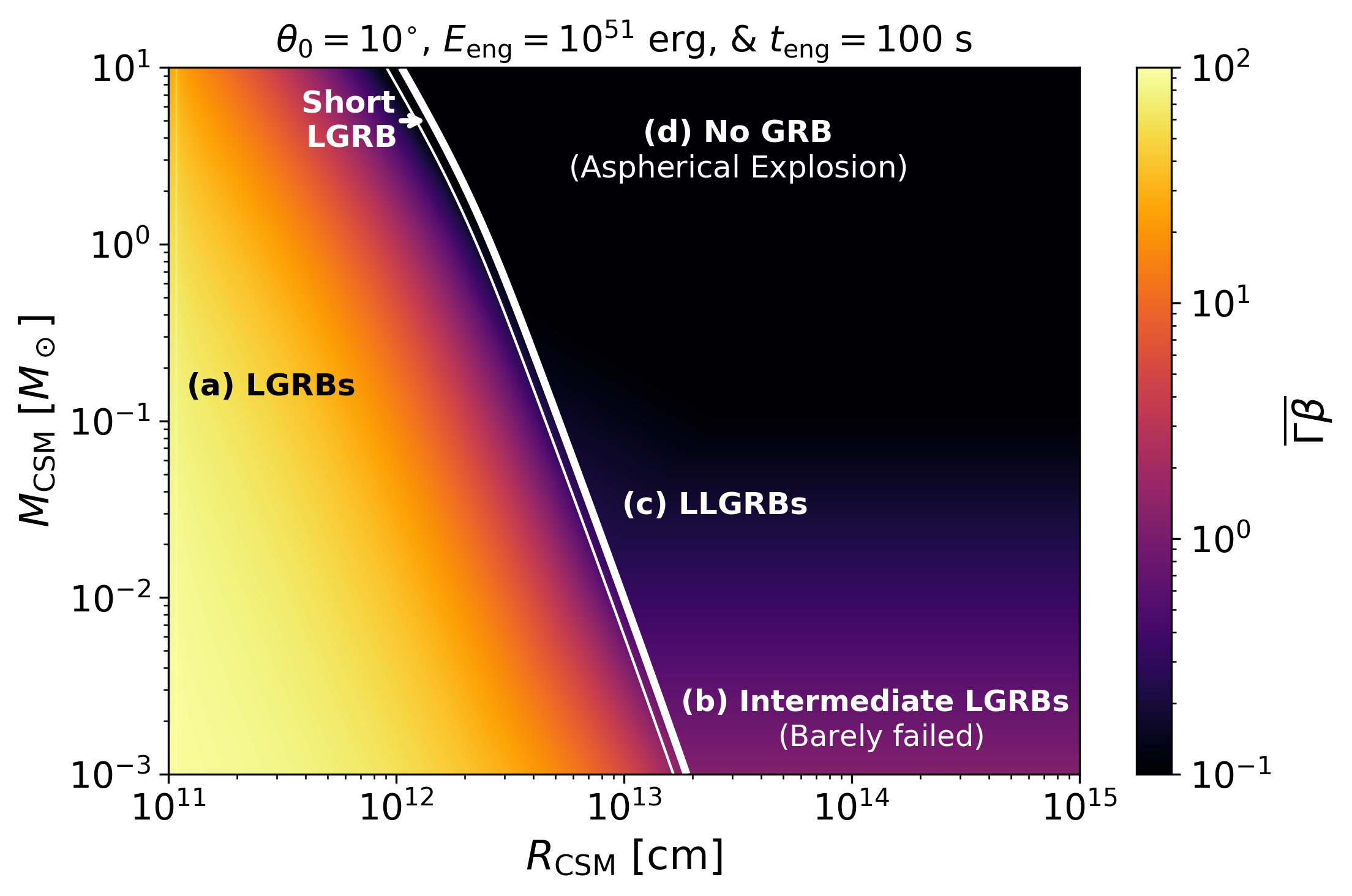} 
  \caption{Same as Figure \ref{fig:phase} for a weak jet ($E_{\rm eng}=10^{51}$ erg).
}
  \label{fig:weak} 
\end{figure}

Figure \ref{fig:weak} shows the same results (as Figure \ref{fig:phase}) for a weak jet case ($E_{\rm eng}=10^{51}$ erg).
Noticeable differences are:
the transition from jet success to jet failure is shifted to lower radii (by a factor of $\sim 10^{0.25}\sim 1.7$; see Equation (\ref{eq:MR simple})); 
the reduced values of $\overline{\Gamma\beta}$ in the failed/barely failed parameter space (because the ratio of energy to mass is lower here);
and the lack of luminous FBOTs as a result of the much lower energy available for cocoon cooling emission (although less luminous FBOTs, similar to the first peak in GRB~060218 could still be produced, as well as afterglow powered non-thermal FBOTs; see \citealt{2025arXiv250316243H,2025arXiv250324266Z}).

These weak jets offer a wider parameter space to explain LLGRBs.
For LLGRBs, follow-up observations targeting FBOT-like thermal emission after the prompt emission are promising, as they potentially offer the opportunity to measure the combination of engine energy, CSM radius, and CSM mass (\citealt{2015ApJ...807..172N}; see \citealt{2025arXiv250316243H} for EP240414a); and ultimately, this could help complete the full picture of their origin and environment (see Section \ref{sec:fbot}).

The weak jet energy here is comparable to that of CCSNe Ibc, which leads to speculation that such weak (and preferably much weaker) jets choked in CSM could theoretically power SN-like emission 
(as discussed in \citealt{2022ApJ...925..148S,2024PASJ...76..863S} for jets in CSM; also see \citealt{2019ApJ...871L..25P,2022MNRAS.517..582E,2023MNRAS.519.1941P} for the possibility of SNe powered by choked jets in the stellar core).

Such jet-driven explosions could produce SNe featuring shock breakout emission, (e.g., such as SN~2008D/XRT~080109; see \citealt{2009ApJ...702..226M}).
A possible signature is morphological asphericity in the SN explosion, which has been observed in Galactic SN remnants (\citealt{2011ApJ...732..114L}; \citealt{2013ApJ...764...50L}) and suggested by observations of extragalactic SNe
\citep[e.g.,][]{2008Sci...319.1220M,2019ApJ...883..120P,2022ApJ...927..180P,2024NatAs...8..111F}.

\subsection{Recent EP events}
In the context of recent events detected by EP,
we speculate that EP240414a might be consistent with a barely failed jet scenario, because of the intermediate intensity of
its prompt emission and afterglow emission (relative to LLGRBs and LGRBs);
and its association with an FBOT  (\citealt{2024arXiv241002315S,2025ApJ...978L..21S,2025ApJ...982L..47V,2024arXiv240919055B}) whose early phase shows thermal emission (\citealt{2025ApJ...982L..47V}).
However, EP250108a (\citealt{2025arXiv250408886E,2025arXiv250408889R,2025arXiv250417516S,2025arXiv250417034L}), with its much softer prompt emission and thermal FBOT emission, appears rather consistent with conventional LLGRBs associated with failed jets.
Considering an extended CSM, we discuss in details the parameter space for EP240414a in \cite{2025arXiv250316243H}, based on its observed FBOT, and the expected cooling and afterglow emissions from the jet-cocoon.


\section{Summary \& Conclusion}
\label{sec:con}
Motivated by growing observational evidence of dense CSM around stripped-envelope SNe, we developed an analytic model to systematically study the dynamics of LGRB jet propagation in a wide range of CSM environments (Figure \ref{fig:picture}).
We considered a conventional LGRB jet with parameters $E_{\rm eng}=10^{52}$ erg, $t_{\rm eng}=10^2$ s, and $\theta_0=10^\circ$, and explored the effects of the CSM radius ($R_{\rm CSM}$) and mass ($M_{\rm CSM}$) on jet evolution.

Starting from the jump conditions [equation (\ref{eq:jump})], we derived the jet head velocity in the CSM [equations (\ref{eq:betah final}) and (\ref{eq:betah analytic})] and obtained a general formula applicable to both Newtonian ($\beta_{\rm h}\ll 1$) and relativistic ($\beta_{\rm h}-1\ll 1$) regimes, as well as intermediate cases (Figure \ref{fig:blend}). 
We present a new refined expression for the breakout time [equations (\ref{eq:tb1}) and (\ref{eq:tb apporx})], which, unlike previous works, includes the factor $1-\beta_{\rm h}$; an important term that has not been considered in previous works focused on jets in massive stars, where $\beta_{\rm h} \ll 1$.
This advancement enables the first systematic analytic treatment of jet propagation across a wide range of CSM properties.

Our analysis shows that jet propagation in extended CSM exhibits fundamental differences from that in massive stars.
Specifically, while jets in dense stellar interiors are characterized by slow head velocities ($\beta_{\rm h}\ll 1$), those in extended CSM can approach $\beta_{\rm h}-1\ll 1$ (see Figures \ref{fig:betah}).
We find that the jet can still be successful even if $t_{\rm b}\gg t_{\rm eng}$ (left panel of Figure \ref{fig:tb}), against the classical success criterion ($t_{\rm b}<t_{\rm eng}$ for successful jets and $t_{\rm b}>t_{\rm eng}$ for failed jets).
To address this, we introduced a criterion based on the jet timescale $t_{\rm jet}$ that reflects the health of the jet as it propagates in a given CSM [in terms of its physical size, equation (\ref{eq:tjet}) and its residual energy, equation (\ref{eq:Ej})].
Consequently, the jet is determined as successful if $t_{\rm b}<t_{\rm eng}/(1-\beta_{\rm h})$; or failed otherwise (Figure \ref{fig:space-time}).
We alternatively expressed this criterion as a function of the CSM/jet parameters and found a strong dependence on $R_{\rm CSM}$ for LGRB jets [equation (\ref{eq:MR simple})].

We examined the evolution of $t_{\rm jet}$ and mapped the parameter space where jets either successfully break out or fail (right panel of Figure \ref{fig:tjet}).
We also evaluated the typical proper velocity $\overline{\Gamma\beta}$ of the jet-cocoon system [equation (\ref{eq:av})].
Between the extremes of (a) successful jets (dominated by relativistic jet material; $\overline{\Gamma\beta} \sim 10-100$) and (c) failed jets (dominated by the outer cocoon; $\overline{\Gamma\beta} \sim 0.1$), we identified (b) an intermediate case where the system is mildly relativistic ($\overline{\Gamma\beta} \sim 1$), driven by a combination of inner and outer cocoons (Figure \ref{fig:jets}).
Following \cite{2022MNRAS.517..582E}, we refer to this case as a ``barely failed" jet.

Our analysis constrains the CSM environments of LGRBs and their observational signatures as follows (Figures \ref{fig:phase} and \ref{fig:weak}) (see Sections \ref{sec:LGRBs}, \ref{sec:LLGRBs}, and \ref{sec:Barely failed GRBs}):

\begin{itemize}
    \item \textbf{(a) LGRBs:} 
    $R_{\rm CSM}^{\rm LGRB} < 10^{13}\text{ cm }(M_{\rm CSM} / 0.1M_\odot)^{-\frac{1}{4}}$ $(E_{\rm eng}/10^{52}\text{ erg})^{\frac{1}{4}}$.
    Such CSM has an insignificant effect on the corresponding conventional jet ($E_{\rm eng}\sim 10^{52}$ erg).
    They exhibit luminous prompt and afterglow emission.
    In this case, non-thermal FBOT-like emission powered by the jet afterglow could be observed, in both on-axis and off-axis jet viewing angles \citep{2025arXiv250324266Z}.
    \item \textbf{(b) Intermediate GRBs} (barely failed jets): 
    $R_{\rm CSM}^{\rm (b)} \gtrsim 10^{13}\text{ cm }(M_{\rm CSM}/0.1M_\odot)^{-\frac{1}{4}}(E_{\rm eng}/10^{52}\text{ erg})^{\frac{1}{4}}$ and $M_{\rm CSM}^{\rm (b)}\lesssim 0.1M_\odot(E_{\rm eng}/10^{52}\text{ erg})$.
    These systems could produce intermediate luminosity prompt emission and afterglows. 
    Thermal FBOT emission powered by the cocoon, or non-thermal FBOT-like emission powered by the jet-cocoon afterglow, could be expected from conventional LGRB jets.
    \item \textbf{(c) LLGRBs:} 
    $R_{\rm CSM}^{\rm LLGRBs} \gtrsim 10^{13}\text{ cm }(E_{\rm eng}/10^{52}\text{ erg})^\frac{1}{4}$ $(M_{\rm CSM}/10^{-1}M_{\odot})^{-\frac{1}{4}}$, $R_{\rm CSM}^{\rm LLGRBs} \lesssim  10^{14} \text{ cm}$, and $M^{\rm LLGRBs}_{\rm CSM} \gtrsim 10^{-1}M_{\odot}(E_{\rm eng}/10^{52}\text{ erg})$.
    These events are characterized by LLGRB-like dim prompt emission and afterglow emissions. They can be produced by weak or conventional LGRB-jets, where in the latter case a potentially luminous thermal FBOT powered by the cocoon cooling emission could be expected.
    \item \textbf{(d) Aspherical explosions with no GRB:} 
    $R_{\rm CSM}\gtrsim 10^{13}$ cm and $M_{\rm CSM}\gtrsim 1M_\odot$. 
    These could result in non-relativistic explosions with $\beta_{\rm c}\sim 0.03\,(M_{\rm CSM}/10M_\odot)^{-\frac{1}{2}}(E_{\rm eng}/10^{52}\text{ erg})^{\frac{1}{2}}$.
\end{itemize}

Our results place constraints on the CSM properties of cosmological LGRBs, assuming $E_{\rm eng}\sim 10^{52}$ erg, and highlight the difference between their CSM and CSM observed in SNe Ibc (see Figure \ref{fig:phase} and Section \ref{sec:1}).
However, because of possible variations in collapsar central engine energies, LLGRBs may arise either from conventional jets in dense CSM or from intrinsically weaker jets (see Section \ref{sec:weak}).
Thus, the higher observed event rates of LLGRBs cannot be solely attributed to a higher abundance of collapsar systems with more CSM presence relative to collapsar systems with less CSM presence.

Finally, we emphasize the importance of thermal FBOT emission in failed and barely failed jets, powered by cocoon cooling emission at large breakout radii.
Such emissions provide a valuable opportunity to infer both the CSM properties and jet characteristics of LLGRBs.
Also, afterglow observations could help distinguish different jet scenarios by constraining their relativistic nature.
Future follow-up observations could significantly improve our understanding of the link between LGRBs, LLGRBs, and their environments.

Recent EP events (EP240414a: \citealt{2024arXiv241002315S,2025ApJ...978L..21S,2025ApJ...982L..47V,2024arXiv240919055B}; and EP250108a \citealt{2025arXiv250408886E,2025arXiv250408889R,2025arXiv250417516S,2025arXiv250417034L}) provide key examples,
where EP240414a appears to be consistent with a barely failed jet (b) (viewed on-axis: \citealt{2025arXiv250316243H}; or slightly off-axis \citealt{2025arXiv250324266Z}); 
EP241021a with its brighter radio afterglow might be consistent with a successful jet (a) \citep{2025arXiv250314588B,2025arXiv250505444G,2025arXiv250507665X,2025arXiv250508781Y}; and
EP250108a may correspond to a completely failed jet (c) \citep{2025arXiv250417516S}.


\section{Data availability}
The data underlying this article will be shared on reasonable request to the corresponding author.

\begin{acknowledgements}
We thank
Katsuaki Asano,  
Sho Fujibayashi,  
Christopher Irwin,
Wataru Ishizaki,  
Shota Kisaka, 
Ryo Sawada,  
Jiro Shimoda,
Akihiro Suzuki,
Kenji Toma,
and Ryo Yamazaki
for their fruitful discussions and comments. 

This research was supported by the Japan Society for the Promotion of Science (JSPS) Grant-in-Aid for Scientific Research No. 23K19059;
No. 22H00130, 23H05430, 23H04900, 23H01172 (K.I.), and No. 24K00668, 23H04899 (K. K.).

Numerical computations were achieved thanks to the following: Cray XC50 of the Center for Computational Astrophysics at the National Astronomical Observatory of Japan, and Cray XC40 at the Yukawa Institute Computer Facility.
\end{acknowledgements}

\bibliography{0-P8}
\bibliographystyle{aasjournal}


\appendix

\section{Approximation analytic expression for $\beta_{\rm h}$}
\label{ap:A}
From equation (\ref{eq:betah final}), we aim to derive an approximated analytic expression for $\beta_{\rm h}$.
From equation (\ref{eq:betah final}), we can deduce that for $0\leqslant  \beta_{\rm h}< 1$ we have $\frac{dC}{d\beta_{\rm h}}>0$.
Hence, for increasing values of $\beta_{\rm h}$, $C(\beta_{\rm h})$ is always growing.
In other words, every value of $C$ gives one unique solution of $\beta_{\rm h}$ (and vice versa).

In the following, we find the asymptotic solutions for equation (\ref{eq:betah final}) at the limits, and use the technique of ``blended asymptotic approximation" to find a general analytic solution by blending the asymptotic solutions numerically in one single analytic form.

\subsection{Asymptotic Solutions}

\subsubsection{Solution for non-relativistic jet head with $C\ll 1$}
At the limit $C \ll 1$ [which physically corresponds to massive/dense ambient media or weak jets; see equation (\ref{eq:c})], we can approximate $\beta_{\rm h}\ll 1$, allowing us to approximate $(1 - \beta_{\rm h})^{-5} \approx 1 + 5\beta_{\rm h} \sim 1$. 
This simplifies the equation (\ref{eq:betah final}) to:
\begin{equation}
\beta_{\rm h} \approx C^{1/3} ,    
\label{eq:c<<1}
\end{equation}
which is the same as the classical form for jet propagation inside massive stars (see \citealt{2011ApJ...739L..55B,2018MNRAS.477.2128H,2021MNRAS.500..627H,2022MNRAS.517.1640G}).

\subsubsection{Solution for relativistic jet heads with $C\gg 1$}

For $C\gg 1$, we can approximate $\beta_{\rm h} \sim 1$ and define $\epsilon = 1 - \beta_{\rm h}$, where $\epsilon \ll 1$. 
Rewriting equation (\ref{eq:betah final}), we obtain
\begin{equation}
\frac{(1 - \epsilon)^3}{\epsilon^5} = C    ,
\end{equation}
and approximating for small $\epsilon$, we get $\frac{1 - 3\epsilon}{\epsilon^5} \approx C$, which gives:
\begin{equation}
\epsilon \approx c^{-1/5}.
\end{equation}
Hence, for $C\gg 1$ we can approximate:
\begin{equation}
\beta_{\rm h} \approx 1 - C^{-1/5}.
\label{eq:c>>1}
\end{equation}

\subsection{Blended Approximation \& analytic form}

From the two solutions above, we can create a smooth transition between the two limits. 
We define a blending function:
\begin{equation}
T(b) = \frac{1}{1 + (C/C_t)^b}    ,
\end{equation}
where $C_t$ controls the transition point, and $b$ controls the sharpness of the transition.

Using this function, we can approximate:
\begin{equation}
\beta_{\rm h} \approx \frac{\left[C^{1/3}+\left(1 - C^{-1/5} \right) {(C/C_0)^b}\right]}{1 + (C/C_t)^b} ,   
\end{equation}
where $C_t$, and $b$ can be found via nonlinear least squares fitting as:

\begin{equation}
\begin{split}
    C_t &\approx 0.6586, \\
    b &\approx 0.7880.
\end{split}
\end{equation}

Thus, the final expression for $\beta_{\rm h}$ is:
\begin{equation}
\beta_{\rm h} \approx \frac{\left[C^{1/3}+\left(1 - C^{-1/5}\right) (C / 0.66)^{0.79} \right]}{1 + (C / 0.66)^{0.79}}      ,
\label{eq:betah approx}
\end{equation}
where $C = \frac{2^5\eta N_{\rm s}^4}{c^3}\frac{L_{\rm {iso,0}}R_{\rm CSM}}{\theta_0^2 M_{\rm CSM}}$ [see equation (\ref{eq:c})].

As shown in Figure \ref{fig:blend}, this function smoothly interpolates between the asymptotic solutions $C\ll 1$ and $C\gg 1$ [equations (\ref{eq:c<<1}) and (\ref{eq:c>>1})] over a very wide physical parameter space (represented by $C$) and gives reasonably accurate values of $\beta_{\rm h}$ (error $< 4\%$).

\section{Glossary of the mathematical symbols}
\label{ap:Glossary}
\begin{table*}
\begin{minipage}{155mm}
\caption{Glossary of the mathematical notations used throughout the paper.}
\begin{tabular}{lll}
    \hline
    \hline
\multicolumn{1}{c}{Notation} & \multicolumn{1}{c}{Definition} & \multicolumn{1}{c}{Range (Reference equations)}\\
    \hline
    \hline
        \multicolumn{3}{c}{Coordinates}  \\
        \hline
        $t_{}$ & Time in the laboratory frame & \\
        $r$ & Radius in the central engine's frame\\
        $\beta$ & Velocity in units of $c$\\
        $\Gamma$ & Lorentz factor & $=(1-\beta^2)^{-\frac{1}{2}}$ \\
        \hline
        \multicolumn{3}{c}{CSM initial parameters}  \\
        \hline
        $R_{\rm CSM}$ & CSM's outer radius & $\sim 10^{11}-10^{16}$ \rm{cm}\\
        $M_{\rm CSM}$ & CSM mass & $\sim 10^{-3}-10^{1}\,M_{\odot}$\\
        $n$ & CSM density profile index & $=2$\\
        $r_0$ & CSM's inner radius ($\sim$ WR radius) & $\sim 4\times10^{10}$ \rm{cm}\\
        \hline
        \multicolumn{3}{c}{Jet initial parameters}  \\
        \hline
        $t_0$ & Jet launch time & $\equiv 0$\\
        $\theta_0$ & Jet initial opening angle & =10$^\circ$\\
        $E_{\rm eng}$ & Total jet energy & $=10^{52}$ \rm{erg} \\
        $t_{\rm{eng}}$ & Jet engine duration & $=10^2$ s\\
        $L_{\rm iso,0}$ & Jet isotropic equivalent luminosity & $\approx \frac{2E_{\rm eng}}{\theta_0^2t_{\rm eng}}\approx\frac{4L_{\rm j}}{\theta_0^2}$\\
        \hline
        \multicolumn{3}{c}{Jet-cocoon dynamics \& breakout}  \\
        \hline
        $\beta_{\rm h}$ & Jet head velocity & (\ref{eq:betah})\\
        $\beta_{\rm t}$ & Jet tail velocity & $\simeq 1$\\
        $t_{\rm b}$ & Breakout time & (\ref{eq:tb1}) \& (\ref{eq:tb apporx})\\
        $t_{\rm jet}$ & Jet's timescale &  (\ref{eq:tjet})\\
        $P_{\rm c}$ & Cocoon pressure & (\ref{eq:S3})\\
        $E_{\rm c}$ ($E_{\rm c,i}$) & Cocoon (internal) energy & (\ref{eq:Pc})\\
        $V_{\rm c}$ & Cocoon volume & (\ref{eq:Vc})\\
        $N_{\rm s}$ & Numerical calibration factor & $\sim 0.5$\\
        $\eta$ & Cocoon internal energy fraction & $\sim1$\\ 
        $E_{\rm j}$ & Energy of the relativistic jet & (\ref{eq:Ej})\\
        $\overline{\Gamma\beta}$ & Typical (energy averaged) proper velocity of the jet-cocoon system & (\ref{eq:av})\\ 
        $\beta_{\rm c}$ ($\Gamma_{\rm c}$) & Cocoon's average velocity (Lorentz factor) & (\ref{eq:Gammac})\\ 
        \hline
        \hline
\end{tabular}
\end{minipage}
\label{tab:ap}
\end{table*}

\label{lastpage}

\listofchanges

\end{document}